\documentstyle[12pt,fleqn,epsfig,espcrc1]{article}

\title{\vspace{-1cm}{\bf Photodisintegration of three- and four- nucleon systems}\thanks{Invited talk at the XVth International Conference on Few-Body Problems in
Physics (22-26 July 1997, Groningen, The Netherlands)}} \author{
  \underline{W. Sandhas}$^{\rm a}$, W. Schadow$^{\rm a}$,
  G. Ellerkmann\address{Physikalisches Institut der Universit\"at Bonn, 
    Endenicher Allee 11-13, D-53115 Bonn, Germany}\thanks{The work of W. S,
  W. Sch., and G. E.  was supported by the Deutsche Forschungsgemeinschaft},
  L.~L.  Howell$^{\rm b}$, and S.~A. Sofianos\address{University of South
    Africa, P.O. Box 392, Pretoria, 0003, South Africa}}

\begin{document}

\maketitle

\abstract{ Three- and four-nucleon photodisintegration processes are quite
  efficiently treated by means of effective two-body integral equations in
  momentum space. We recall some aspects of their derivation, present
  previous and most recent results obtained within this framework, and discuss
  general features, trends and effects observed in these investigations:  At
  low energies final-state interaction plays an important role. Even more
  pronounced is the effect of meson exchange currents.  A considerable
  potential dependence shows up in the low-energy peak region. The different
  peak heights are found to be
  closely correlated with the corresponding binding energies. Above the
  peak region only the difference between potentials with or without $p$-wave
  contributions remains relevant. In the differential cross sections the
  electric quadrupole contributions have to be taken into account.  The
  remarkable agreement between theory and experiment in $p$-$d$ radiative
  capture is achieved only when incorporating this contribution, 
  together with most of the
  above-mentioned effects. In the final part of this report we briefly review
  also methods developed, and results achieved in three- and four- nucleon
  electrodisintegration. We, in particular, compare them with a recent access
  to this problem, based on the construction of nucleon-nucleus potentials via
  Marchenko inversion theory.}

\section {THREE-BODY PROCESSES}

Starting form the momentum space formulation of three-nucleon collision
processes, we sketch the transition to the corresponding photodisintegration
equations. Results obtained within this approach are presented, general trends,
sensitivities, and correlations are discussed.

\subsection{Deuteron photodisintegration}

The characteristic features of the momentum-space approach are most easily
described in the two-body case. There, the scattering amplitude is given by
the on-shell restriction $ (p^{\prime 2} = p^{2})$ of the $T$-matrix

\setcounter{equation}{0}

\begin{equation}
\label{eqtdef}
T(\vec p^{\,\prime},\vec p \,) = \langle{\vec p} ^{\, \prime} |T| {\vec p} \,\rangle  = {^{(-)}}{\langle} 
{\vec p}^{\, \prime} |V| {\vec p} \,\rangle \,,
\end{equation}

\noindent
where $|{\vec p}\, \rangle$ represents the incoming plane-wave state, and
$^{(-)}\langle {\vec p}^{\, \prime}| $ the outgoing scattering state.  The
T-operator in (\ref{eqtdef}) satisfies the Lippmann-Schwinger (LS) equation

\begin{equation}
\label{eqtmat}
T = V + V \, G_0 \, T \,.
\end{equation}

\noindent
Being closely related to the physical amplitude, this operator is well-
defined in momentum space also off the energy shell $(p^{\, \prime 2} \neq
p^2)$, and the same holds true for the momentum representation
of (\ref{eqtmat}),

\begin{equation}
\label{eqtmatint}
T ({\vec p}^{\, \prime}, {\vec p}\,) = V ( {\vec p}^{\, \prime}, {\vec p}\,) + 
\int \! d^3
p^{\, \prime \prime} \, V ({\vec p}^{\, \prime}, 
{\vec p}^{\, \prime \prime}) \,
G_0 (p^{\, \prime \prime}) \, 
T ({\vec p}^{\, \prime \prime}, {\vec p}\,) \,.
\end{equation}

\noindent
There is only a singularity in the Green function $G_0 (p^{\, \prime \prime })
= (p^2 / 2 \mu + i \epsilon - p^{\, \prime \prime 2} / 2 \mu)^{-1}$ at the
on-shell point $p^{\, \prime \prime 2} = p^2$, which however is easily taken
care of. A useful way of doing this is provided by the $K$-matrix approach or,
more efficiently, by the $W$-matrix method. We mention this method since it
yields a particularly convenient representation of the two-body input to the
three- and four-body treatments discussed in the following.

Analogously to (\ref{eqtdef}), the amplitude for the photodisintegration of
the deuteron $|\psi_d \rangle$ is given by

\begin{equation}
\label{eqphotodef}
M ({\vec p}\,) = {^{(-)}}{\langle} {\vec p} \, |H_{em}| \psi_d \rangle
= \langle {\vec p} \, | (1 + T G_0) H_{em} | \psi_d \rangle \,,
\end{equation}

\noindent
where $H_{em}$ denotes the electromagnetic operator. Here we have made use of
the fact that the momentum state $\langle \vec p \, |$ is mapped upon the
scattering state $^{(-)}\langle \vec p \,|$ via the adjoint M{\o}ller operator
$(1 + T \, G_{0})$. This representation shows that the amplitude
(\ref{eqphotodef}) is composed of a plane wave (Born) term

\begin{equation}
\label{eqborn}
B ({\vec p}\,) = \langle {\vec p}\, |H_{em}| \psi_d \rangle
\end{equation}

\noindent
and a further contribution $\langle {\vec p}\, | T G_0 H_{em}| \psi_d \rangle$
which, via $T$, takes into account the final-state interaction(FSI).
Inserting in this expression the LS equation (\ref{eqtmat}), we obtain for
$M({\vec p}\,)$ the integral equation

\begin{equation}
\label{eqphotoint}
M ({\vec p}\,) = B({\vec p}\,) + \int \! d^3 {p}^{\, \prime}\, V 
({\vec p},{\vec p}^{\, \prime}) \,
G_0 ({\vec p}^{\, \prime}) M ({\vec p}^{\, \prime}) \,.
\end{equation}

\noindent
Equations (\ref{eqtmatint}) and (\ref{eqphotoint}) are evidently of the same
structure. Their kernels $V \, G_0$, i.e., their most relevant ingredients,
are identical.  But, the nuclear Born approximation $V(\vec p, \vec p^{\,
  \prime}) $ in the inhomogeneity of (\ref{eqtmatint}) is replaced in
(\ref{eqphotoint}) by the photonuclear Born term (\ref{eqborn}). In other
words, with this comparatively simple replacement, any working NN collision
program based on (\ref{eqtmatint}) can immediately be applied to deuteron
photodisintegration.

The present derivation serves primarily to exhibit structural aspects, which
are typical also for the corresponding three- and four-nucleon
generalizations.  It, moreover, provides a quite efficient numerical tool,
characterized by considerable stability, as emphasized in the context of
(\ref{eqtmatint}). But, of course, in the two-body case there are numerous
alternative techniques at our disposal (for a comprehensive review see
\cite{Aren91}). In the three- or four-body case, however, the generalization
of (\ref{eqphotoint}) represents one of the most efficient standard
techniques. 

In electric dipole (E1) approximation the amplitude (\ref{eqphotodef}) is
given by

\begin{equation}
\label{eqphoto-e1}
M ({\vec p}\,) = - \frac{e}{2 \mu} \, {^{(-)}}{\langle} 
{\vec p}\, |\hat 
{\epsilon} \cdot {\vec P} | \psi_d \rangle  =
- \frac{ie}{2} \, {^{(-)}}{\langle} {\vec p}\, |\hat {\epsilon}
 \cdot [H_0,{\vec X}] |\psi_d \rangle \,,
\end{equation}

\noindent
where $\hat {\epsilon}$ denotes the polarization vector of the incident
photon. Replacing in the commutator the free Hamiltonian $H_0 = P^2 / 2 \mu$
by the full Hamiltonian $H = H_0 + V$, this expression goes over into

\begin{equation}
\label{eqphotomec}
M ({\vec p}\,) = - \frac{ie}{2} \, (E_f - E_i) \, {^{(-)}}{\langle} 
{\vec p}\, |\hat {\epsilon} \cdot {\vec X}| \psi_d \rangle \,.
\end{equation}

\noindent
with $(E_f - E_i) = E_ {\gamma}$ being the difference of the final energy
$E_{f} = p^2/ \mu$ and the initial deuteron energy $E_{i} = E_d$, i.e., the
energy $E_{\gamma}$ of the photon.

\noindent
For local potentials, which commute with ${\vec X}$, the transition to
(\ref{eqphotomec}) is only another, in some approaches technically more
convenient way of writing the amplitude (\ref{eqphoto-e1}). For the typically
non-local nuclear potentials, the replacement of (\ref{eqphoto-e1}) by
(\ref{eqphotomec}) leads in general to quite different results.  But,
according to Siegert's theorem \cite{Sieg37}, it is just this replacement,
which takes into account the effect of meson exchange currents (MEC) in the
nucleus.  Siegert's argumentation involves, of course, an approximation. But,
at least in the deuteron case it has been shown that the results obtained in
this way fully agree with calculations based on a Hamiltonian which includes
the MEC contributions explicitly \cite{Aren91}. In the three- and four-body
cases the validity of Siegert's theorem is usually taken for granted, an
assumption which would need a similar verification as in the two-body case.

\begin{figure}[hbt]
\begin{minipage}[t]{75mm}  
  \psfig{file=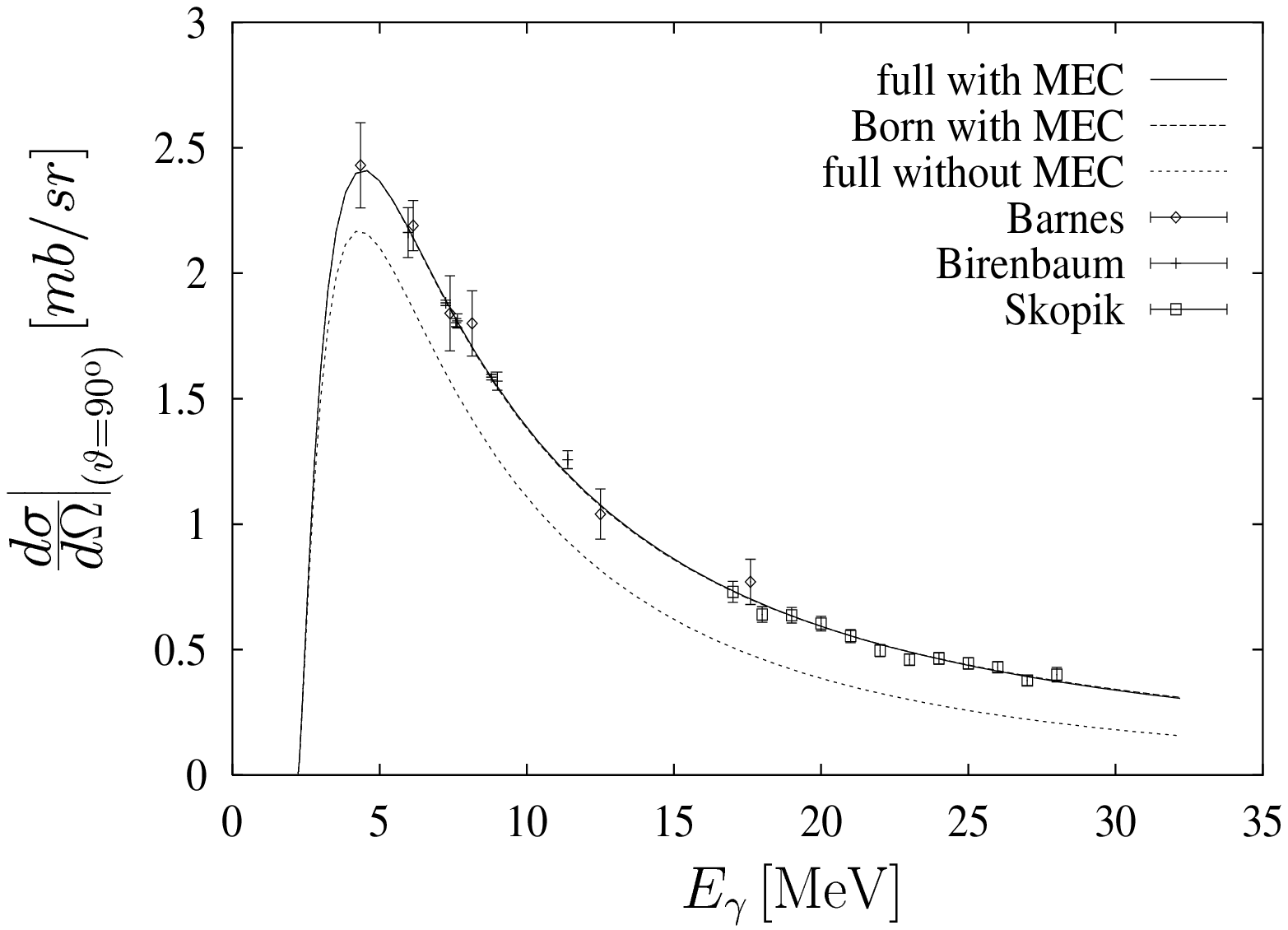,width=49mm} \vspace{-5mm}
\caption{\label{figdeut1}
  Role of meson exchange currents and final-state interaction. The
  experimental data are from  \protect{\cite{Barn52,Biren85,Skop74}}.}
\end{minipage} 
\hspace{5mm}
\begin{minipage}[t]{75mm}
  \psfig{file=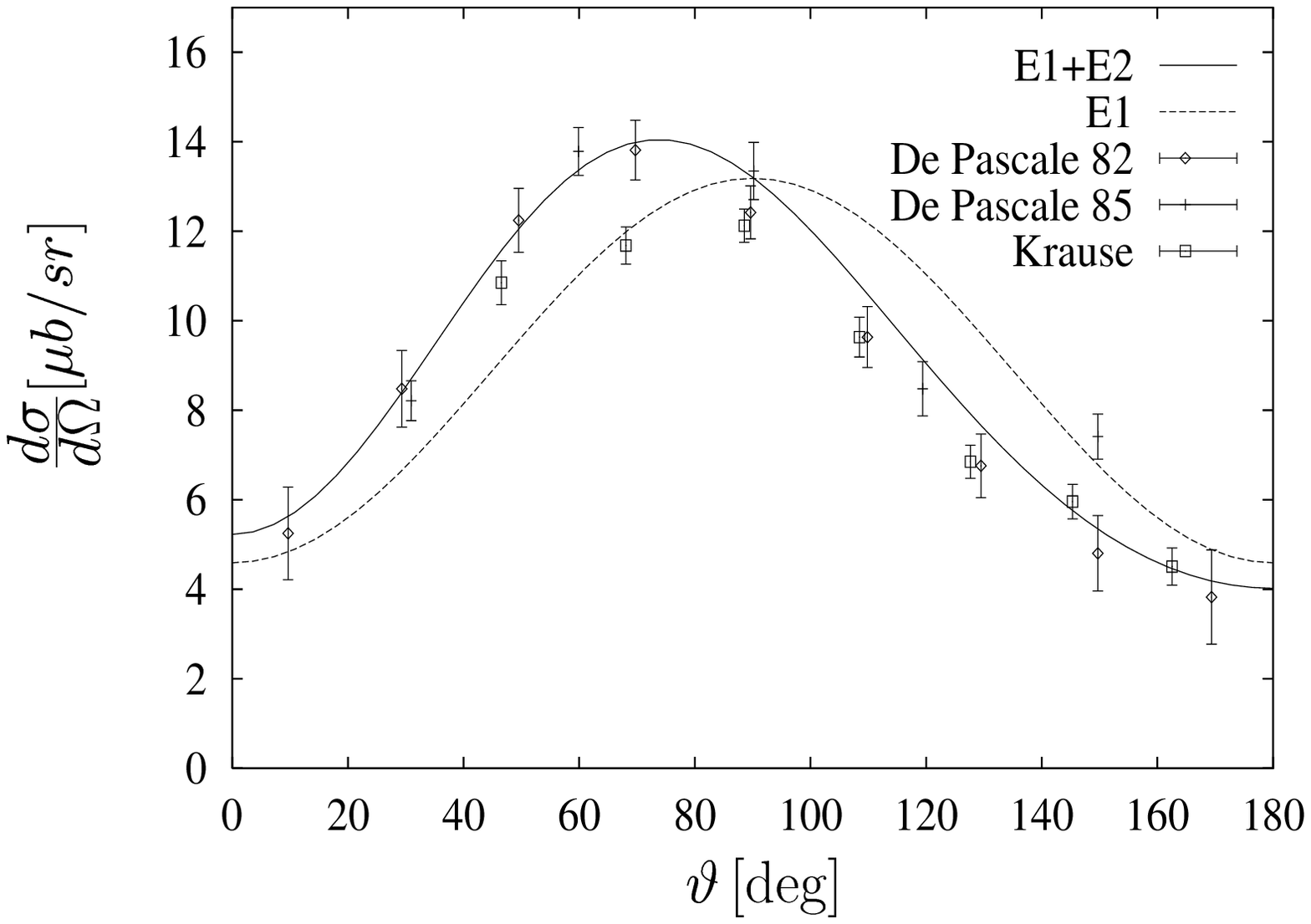,width=49mm} \vspace{-5mm}
\caption{\label{figdeut2} Differential cross section for
  deuteron photodisintegration with the Bonn {\sl B} potential in $E1$ and $E1$
  + $E2$ \protect{approximation} at $E_\gamma$ = 60 MeV. The experimental data are 
  from \protect{\cite{Depas82,Krause88}.}}
\end{minipage}
\end{figure}

Fig. \ref{figdeut1} shows the differential cross section at $\vartheta = 90^0$
in $E1$ approximation, obtained for the Paris potential by solving
(\ref{eqphotoint}).  The full solution is almost identical with the Born
result, i.e., the FSI is practically negligible in the energy region
considered. But there is a clear MEC effect. Fig.  \ref{figdeut2} shows the
differential cross section in $E1$ approximation,  and with inclusion of the
electric quadrupole $(E2)$ contribution.  It should be mentioned that these
results are in complete agreement with an independent solution in Ref.
\cite{Schwambprivate}.

One of the main questions being asked in the following is, whether similar or
quite different FSI, MEC or $E2$ effects occur in the corresponding three- and
four-nucleon processes.

\subsection {Three-body equations} 

In the three-body case the $T$-matrix for rearrangement processes $\alpha +
(\beta \gamma) \rightarrow \beta + (\alpha \gamma)$ is given by

\begin{equation}
\label{eq3btmat}
T_{\beta m, \alpha n} ({\vec q}^{\, \prime}_{\beta}, {\vec q}_{\alpha}) = 
\langle {\vec q}^{\, \prime}_{\beta} | \langle
 \psi _{\beta m} | U_{\beta \alpha} |
\psi_{\alpha n} \rangle | {\vec q}_{\alpha} \rangle \,,
\end{equation}

\noindent
where ${\vec q}_{\alpha}$ denotes the momentum of particle $\alpha$ relative
to the bound state $| \psi _{\alpha n} \rangle$ of particles $\beta$ and
$\gamma$. Thus, due to the various initial and final configurations ($\alpha,
\beta$ = 1, 2 or 3), we have, instead of a single $T$-operator, a set of $3
\times 3$ transition operators. Correspondingly, the single LS operator
identity (\ref{eqtmat}) is replaced by the coupled set of Faddeev-type
Alt-Grassberger-Sandhas (AGS) equations \cite{Alt67}

\begin{equation}
\label{eqags}
U_{\beta \alpha} = (1 - \delta _{\beta \alpha}) G^{-1}_0 + \sum
\limits_{\gamma \neq \beta} \, T_{\gamma} G_0 \,U_{\gamma \alpha} \,.
\end{equation}

\noindent
These equations contain in their kernel the three subsystem $T$-operators
$T_{\gamma} = T_{\alpha \beta}$. In other words, they represent in a most
condensed way the relation between two-body input and three-body output.  Of
particular importance is that, like (\ref{eqtmat}), they are well-defined in
momentum space, and thus amenable to the particularly accurate momentum- space
methods.

As three-body operators, the $U_{\beta\alpha}$ act on the momentum state $|
{\vec q}_{\alpha} \rangle$ of particle $\alpha$ relative to the
$(\beta\gamma)$ system, and on the momentum state $|{\vec p}_{\alpha}\rangle$
within this subsystem. In momentum representation the AGS equations, hence,
are two-dimensional after partial wave decomposition. They were used
successfully in this direct form which, however, needs a considerable
numerical effort.  From the very beginning applications of (\ref{eqags}),
therefore, made use of the most typical aspect of Faddeev-type equations, the
occurrence of $T_{\gamma}$ in their kernel.  Representing these operators in
separable form,

\begin{equation}
\label{eqtmatsep}
T_{\gamma} = \sum\limits_{n\, m} | \chi_{\gamma n} \rangle \,t_{\gamma,nm} 
\,\langle \overline{\chi}_{\gamma m} | \,,
\end{equation}

\noindent
Eq. (\ref{eqags}) reduces to a set of effective two-body equations of LS form
which, in matrix notation, reads

\begin{equation}
\label{eq3btmatint}
{\cal T} ({\vec q}^{\, \prime},{\vec q}\,) = {\cal V} ({\vec q}^{\, \prime},
{\vec q}\,)
+ \int \! d^3 q^{\, \prime \prime} \, {\cal V} ({\vec q}^{\, \prime} \,
{\vec q}^{\, \prime \prime}) \, {\cal G}_0 (q^{\, \prime \prime}) \,
 {\cal T} ({\vec q}^{\, \prime \prime} {\vec q}\,) \,.
\end{equation}

\noindent
Here, the matrix elements of ${\cal T}$ are off-shell extensions of
(\ref{eq3btmat}), the elements of ${\cal V}$ are exchange potentials

\begin{equation}
{\cal V}_{\beta m, \alpha n} ({\vec q}^{\, \prime}_{\beta}, {\vec q}_{\alpha})
= (1 - \delta_{\beta\alpha}) \, \langle {\vec q}^{\, \prime}_{\beta} | 
\langle \overline{\chi}_{\beta m} | G_0 | \chi_{\alpha n} \rangle 
|{\vec q}_{\alpha} \rangle
\end{equation}

\noindent
and the free Green function is given by

\begin{equation}
{\cal G}_{0; \beta m, \alpha n} = \delta_{\beta \alpha} \, t_{\alpha, n m} \,.
\end{equation}

\noindent
After this reduction, the resulting matrix LS equation (\ref{eq3btmatint}) is,
of course, to be antisymmetrized.  The analogy of (\ref{eqtmatint}) and
(\ref{eq3btmatint}) indicates that also the photodisintegration equation
(\ref{eqphotoint}) has its counterpart in the present case. A procedure,
closely related to the steps leading to (\ref{eqphotoint}), in fact, yields
for the two-fragment photodisintegration of the triton, $\gamma + t
\rightarrow n+d$, the effective two-body equation

\begin{equation}
\label{eq3bphotoint}
{\cal M} ({\vec q}\,) = {\cal B} ({\vec q}\,) + \int \! d^3 q^{\, \prime} \,
{\cal V} ({\vec q}, {\vec q}^{\, \prime})\, {\cal G}_0 ({\vec q}^{\, \prime}) \,
{\cal M} ({\vec q}^{\, \prime}) \,.
\end{equation}

\noindent
Here, ${\cal M} ({\vec q}\,)$ and ${\cal B} ({\vec q}\,)$ are off-shell
extensions of the full photodisintegration amplitude

\begin{equation}
\label{eq3bphotoampl}
M ({\vec q}\,) = {^{(-)}}{\langle} {\vec q}; \psi_d |H_{em}| \psi_t \rangle 
\end{equation}

\noindent
and of the corresponding plane wave (Born) approximation

\begin{equation}
\label{eq3bborn}
B ({\vec q}\,) = \langle {\vec q} \,|\langle \psi_d| H_{em}| \psi_t \rangle \,,
\end{equation}

\noindent
with ${\vec q}$ the relative momentum between the outgoing nucleon and the
deuteron, and $| \psi_t \rangle$ the triton wave function.

The decisive advantage of this formalism is that the appropriate construction
of ${\cal V}$ and ${\cal G}_0$, including tests of accuracy, is well known
from the purely nuclear case, and has not to be repeated. What remains to be
done, when going over to the photonuclear equation (\ref{eq3bphotoint}), is
the consistent determination of the off-shell Born term ${\cal B}({\vec
  q}\,)$.  For details we refer to the papers quoted in the next section.

\medskip

Let us make some remarks on the derivation of (\ref{eq3bphotoint}). In
Eq.(\ref{eqphotodef}) it is the adjoint M{\o}ller operator $(1 + T G_0)$ which
maps $\langle {\vec p}\,|$ onto $^{(-)}{\langle} {\vec p} \, |$. Analogously,
we have in the present case

\begin{equation}
M ({\vec q}_{\beta}) = \langle {\vec q}_{\beta} | \langle \psi_d | 
(\delta_{\beta \alpha} + U_{\beta a} G_{\alpha}) H_{em} | \psi_t \rangle\,,
\end{equation}

\noindent
with $G_{\alpha} = (E_{f} + i \epsilon- H_0 - V_{\alpha})^{-1}$ the channel
resolvent of the $\alpha = (\beta \gamma)$ subsystem.  Proceeding in strict
analogy to the derivation of (\ref{eqphotoint}), i.e., inserting here, instead
of the LS equation, the AGS equations with $T_{\gamma}$ being chosen according
to (\ref{eqtmatsep}), we end up with an effective two-body equation of the
form (\ref{eq3bphotoint}) with the matrix elements of the inhomogeneity being
given by

\begin{equation}
\label{eq3bborn2}
{\cal B}_n ({\vec q}_{\beta}) = 
\langle {\vec q}_{\beta} | \langle \chi_{\beta m} | G_0\, H_{em} | \psi_t \rangle \,.
\end{equation}

\noindent
We, moreover, mention that consistently with the approximation
(\ref{eqtmatsep}) the triton wave function may be determined by solving the
homogeneous counterpart of equation (\ref{eq3btmatint}).  For details we refer
to \cite{Schad97a,Cant97a,SchadowHaidtobepublished}.

$\!$Three-body photonuclear integral equations were first derived within the
Faddeev framework by Barbour and Phillips \cite{Barb67}. A rather transparent
derivation, based on the AGS equations, and thus more closely related to the
above argumentation, was performed by Gibson and Lehman \cite{Gibs75}. It is,
in fact, their approach which has been taken over in subsequent, more
sophisticated applications or extensions of momentum-space techniques in
photonuclear few-body physics. Only in parenthesis we mention that a
derivation, which exhibits in great detail the structural equivalences pointed
out above, is found in \cite{Sandhas86a}.

In $E1$ approximation the electromagnetic operator $H_{\rm em}$ in the above
three-nucleon relations is given by

\begin{equation}
\label{eqhem1}
H^{\, \prime}_{E1} = - \frac{e}{m} \sum\limits^3_{j=1} \frac{1+\tau^j_z}{2}
\, \hat {\epsilon} \cdot {\vec P}_j \,.
\end{equation}

\noindent
With inclusion of meson exchange currents (MEC) we have, according to
Siegert's theorem, an expression analogous to (\ref{eqphoto-e1}),

\begin{equation}
\label{eq3bmmec}
M ({\vec q}\,) = - ie \, (E_f - E_i)\, {^{(-)}}{\langle} {\vec q} ;
\psi_d | \sum\limits^3_{j=1} \frac{1 + \tau^j_z}{2} \, \hat
{\epsilon} \cdot {\vec X}_j | \psi_t \rangle \,.
\end{equation}

\noindent
Here ${\vec P}_j, {\vec X}_j$ and $\tau^j_z$ are the momentum and position
operators, and the isospin Pauli matrices of the three nucleons involved.  As
in (\ref{eqphotomec}) the difference of the final energy $E_f = q^2/2M + E_d$
and the initial triton energy $E_i = E_t$ is just the energy of the incident
photon, $(E_f - E_i) = E_\gamma$. Let us mention that for $s$-wave potentials
only that part of the $E1$ electromagnetic operator which acts between the
fragments survives, while, due to its $p$-wave structure, the contribution to
$H_{E1}$ acting within the deuteron vanishes.  The electromagnetic operator
then reduces to

\begin{equation}
\label{eqhem1yama}
H_{E1} =
- ie \, (E_f - E_i) \left( \frac{\tau^1_z + \tau^2_z}{2} - \tau^3_z \right) 
\hat{\epsilon} \cdot {\vec R}_3 \,,
\end{equation}

\noindent
where ${\vec R}_3$ denotes the relative position operator between the
elementary particle 3 and the bound state of particles 1 and 2. Although, in
general, higher partial waves are quite relevant in realistic potentials,
(\ref{eqhem1yama}) remains the dominant part of $H_{\rm em}$ also in such
cases.

In the following considerations of the total cross section, the $E1$
contributions are sufficient. A considerable influence of the electric
quadrupole $(E2)$ contribution, however, shows up in the differential cross
section.  The corresponding operator is given by

\begin{equation}
\label{eqhem2}
H^{\, \prime}_{E2} = - \frac{ie}{2m} \sum\limits^3_{j=1}
\frac{1 + \tau^j_z}{2} \, (\hat{\epsilon} \cdot {\vec P}_j \;
{\vec k}_{\gamma} \cdot {\vec X}_j +\hat{\epsilon}
\cdot {\vec X}_j \; {\vec k}_{\gamma} \cdot {\vec P}_j) \,,
\end{equation}

\noindent
with ${\vec k}_{\gamma}$ the momentum of the incident photon. As in the above
relations MEC are included, according to  Siegert's theorem, via the replacement

\begin{equation}
\label{eqhem2mec}
H_{E2} = \frac{e}{2} \,( E_f - E_i ) \sum\limits^3_{j=1}
\frac{1 + \tau^j_z}{2} \, \hat{\epsilon} \cdot {\vec X}_j \;
{\vec k}_{\gamma} \cdot {\vec X}_j \,.
\end{equation}

\subsection{Three-nucleon photodisintegration results}

First applications of the above formalism were performed in \cite{Barb67} and
\cite{Gibs75}.  However, only the latter calculations employed the same
separable Yamaguchi potential for the determination of ${\cal V} \, {\cal
  G}_0$ and the triton wave function in ${\cal B} (\vec q\,)$. Apart from a
physically motivated different choice of the potential parameters in the
bound-state case, these calculations, hence, represent the first consistent
treatment of the problem within a simple separable potential model.  Let us
recall in this context that the Yamaguchi potential $V^s_{l=0} (p^{\, \prime},
p) = \chi (p^{\, \prime}) \, \lambda \,\chi(p)$, with $\chi (p) =
(p^2+\beta^2)^{-1}$, is a pure $s$-wave potential, of course with different
choices of the parameters $\lambda$ and $\beta$ in the triplet and singlet
channels, leading thus to an $s$-wave rank-one separable $T$-matrix.

\begin{figure}[htb]
  \centerline{ \psfig{file=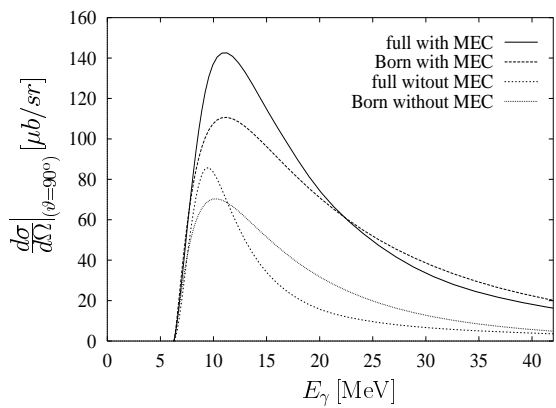,width=49mm}
    \phantom{000000000000000000}} \vspace{-5mm}
\caption{\label{figtritonmec}Role of meson exchange currents in the full
  solution and in Born approximation for the Paris potential ($j \leq
  1^{+}$).}
\end{figure}

It is well known that, despite the simplicity of this model, the resulting
$n$-$d$ cross sections reflect essential aspects of the three-nucleon
collision problem fairly well, and the same holds true for the corresponding
photodisintegration results of \cite{Barb67} and, in particular, of
\cite{Gibs75}. The main observation made already in these early solutions of
the full equations (\ref{eq3bphotoint}) is an about 25 \% enhancement of the
cross section in the low-energy peak region as compared to the Born
approximation.  This feature appears to be independent of the potential
employed \cite{Schad97a}.  Figure \ref{figtritonmec} shows this 
enhancement for the Paris potential. In view of the fact that the
outgoing deuteron is a loosely bound, easily polarizable system, such a strong
FSI effect is not surprising, while in the elementary-particle NN collision it
was found to be negligible (see Sec.~1.1). Even more pronounced (about a factor
of 1.7) is the effect of meson exchange currents.  It is also much stronger
than in the two-body case, an observation easily referred to the higher number
of nucleons involved.

The transition operator $U_{0 \alpha}$ in the $nd \rightarrow nnp$ break-up
  amplitude 
  $\langle {\vec q}^{\,   \prime}, {\vec p}^{\, \prime} | U_{0 \alpha} |
  \psi_d \rangle | {\vec q}\, \rangle$   is obtained by putting
$\beta = 0$ in the AGS equations (\ref{eqags}) \cite{Alt67}. This, however, 
means that it  appears as a sum over the rearrangement
operators $U_{\beta \alpha}$, and the same holds true for the corresponding
amplitudes in the effective two-body formulation, both of the purely nuclear
and the corresponding photonuclear processes. For details concerning the
$\gamma \,t \rightarrow nnp$ formalism we refer to \cite{Gibs76b}.
Calculations along this line have been performed exclusively for the Yamaguchi
potential in \cite{Barb67} and \cite{Gibs76b}. Again the more consistent
calculations of \cite{Gibs76b} are in better agreement with experiment,
although clear discrepancies are also evident.  This calls for a repetition of
these calculations with more realistic potentials.

\subsection{Potential dependence}

One of the main questions in few-body physics is the sensitivity of relevant
observables to the underlying two-body potentials. Due to the fact that the
photodisintegration amplitudes contain both the three-body bound and continuum
states, these processes are expected to be a good testing ground in this
respect. It, in fact, turns out that there are considerable differences
between the photonuclear results obtained for the Yamaguchi, Malfliet- Tjon,
Paris, Bonn {\sl A} and Bonn {\sl B} potentials.

To represent these potentials or the corresponding $T$-matrices in separable
form, various techniques have been developed and tested in the purely nuclear
case. In the photodisintegration problem it were essentially the $W$-matrix
method \cite{Bart86} and the Ernst-Shakin-Thaler (EST) expansion \cite{Erns73}
which have been employed within the above momentum-space formalism. The
$W$-matrix method is characterized by a particularly simple separable
approximation of the $T$-matrix, which preserves all its main analytical and
unitary properties.  The EST expansions of the Paris, Bonn {\sl A} and {\sl B}
potentials developed by the Graz group (PEST, BAEST and BBEST)
\cite{Haid84,Haid86a}, have led to the first fully reliable realistic results
in the three-nucleon problem \cite{Haid86b,Koike87a}.  In the following
applications to photoprocess an improved parametrization by Haidenbauer is
used \cite{Haidpriv}.

\begin{figure}[hbt]
\begin{minipage}[t]{7.5cm}  
  \psfig{file=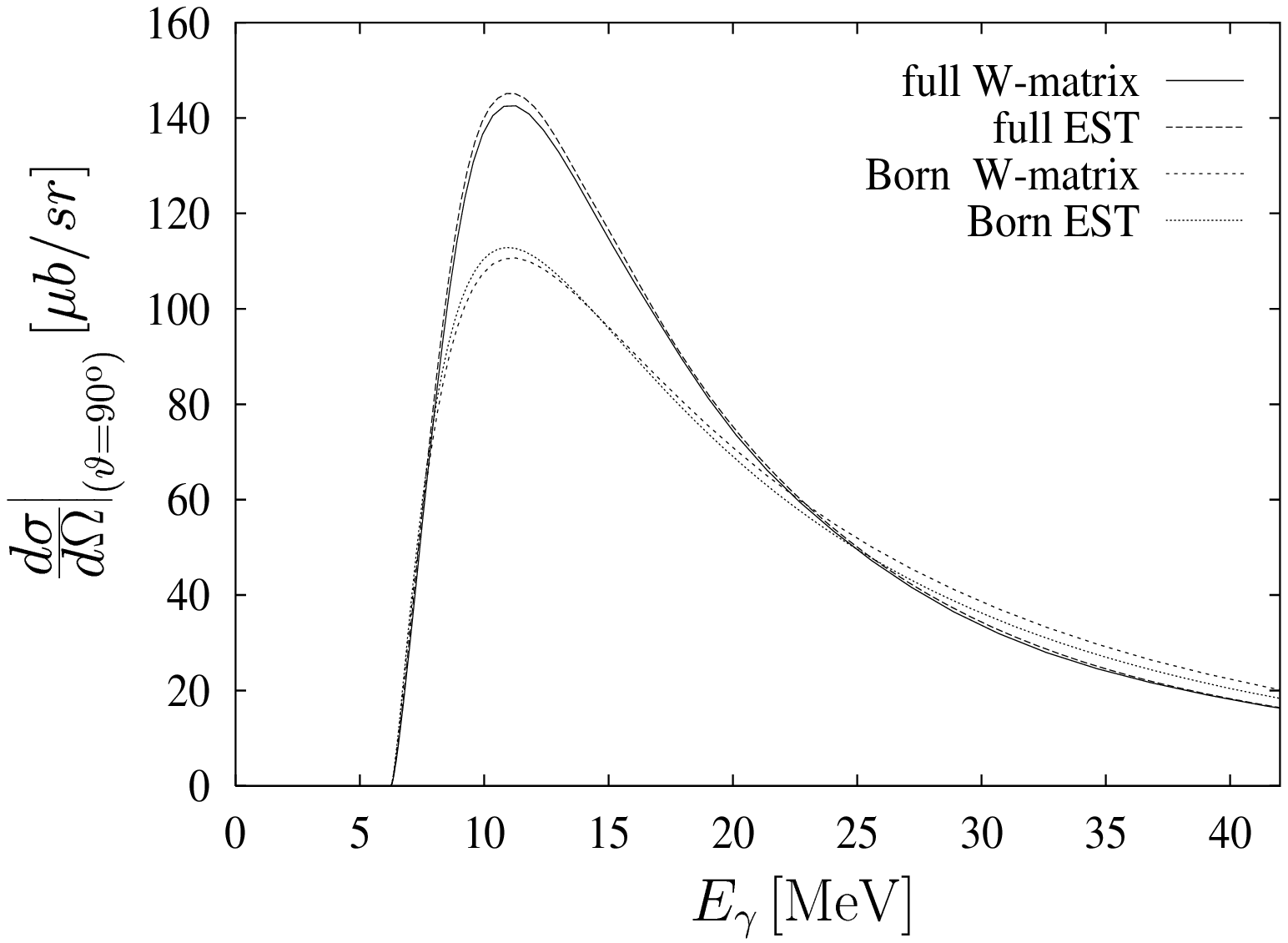,width=49mm} \vspace{-5mm}
\caption{\label{figpape} Comparison of the cross sections obtained for the
  Paris potential $(j \leq 1^{+})$ using the $W$-matrix and the EST method.}
\end{minipage} 
\hspace{5mm}
\begin{minipage}[t]{7.5cm}
  \psfig{file=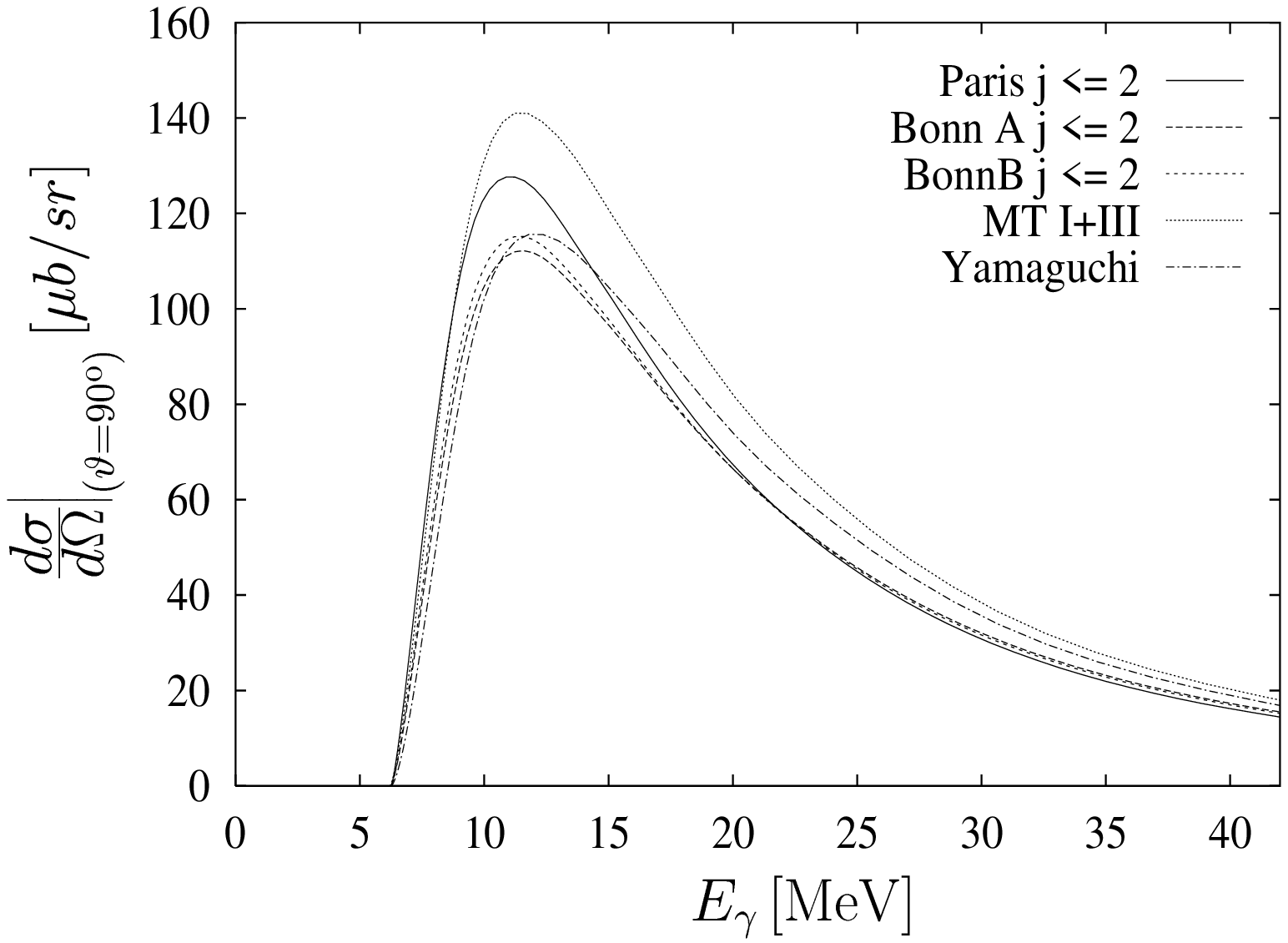,width=49mm} \vspace{-5mm}
\caption{\label{figpotcomp}Cross sections for different potentials}
\end{minipage}
\end{figure}

Fig. \ref{figpape} gives a good indication for the agreement between both
techniques, although for a somewhat limited positive-parity two-body input $(j
\leq 1^+)$. The results given below for the Malfliet-Tjon potential are based
on the $W$-matrix method. For the Paris and Bonn potentials, $W$-matrix results
have also been published \cite{Schad97a}.  The number of subsystem partial
waves employed in these calculations, however, turned out to be too small. We
therefore present, instead of these previous results, most recent fully
converged calculations \cite{Schadowcontribution,Schadowtobepublished}
for the PEST, BAEST, and BBEST potentials.

\begin{figure}[hbt]
\begin{minipage}[t]{7.5cm}  
  \psfig{file=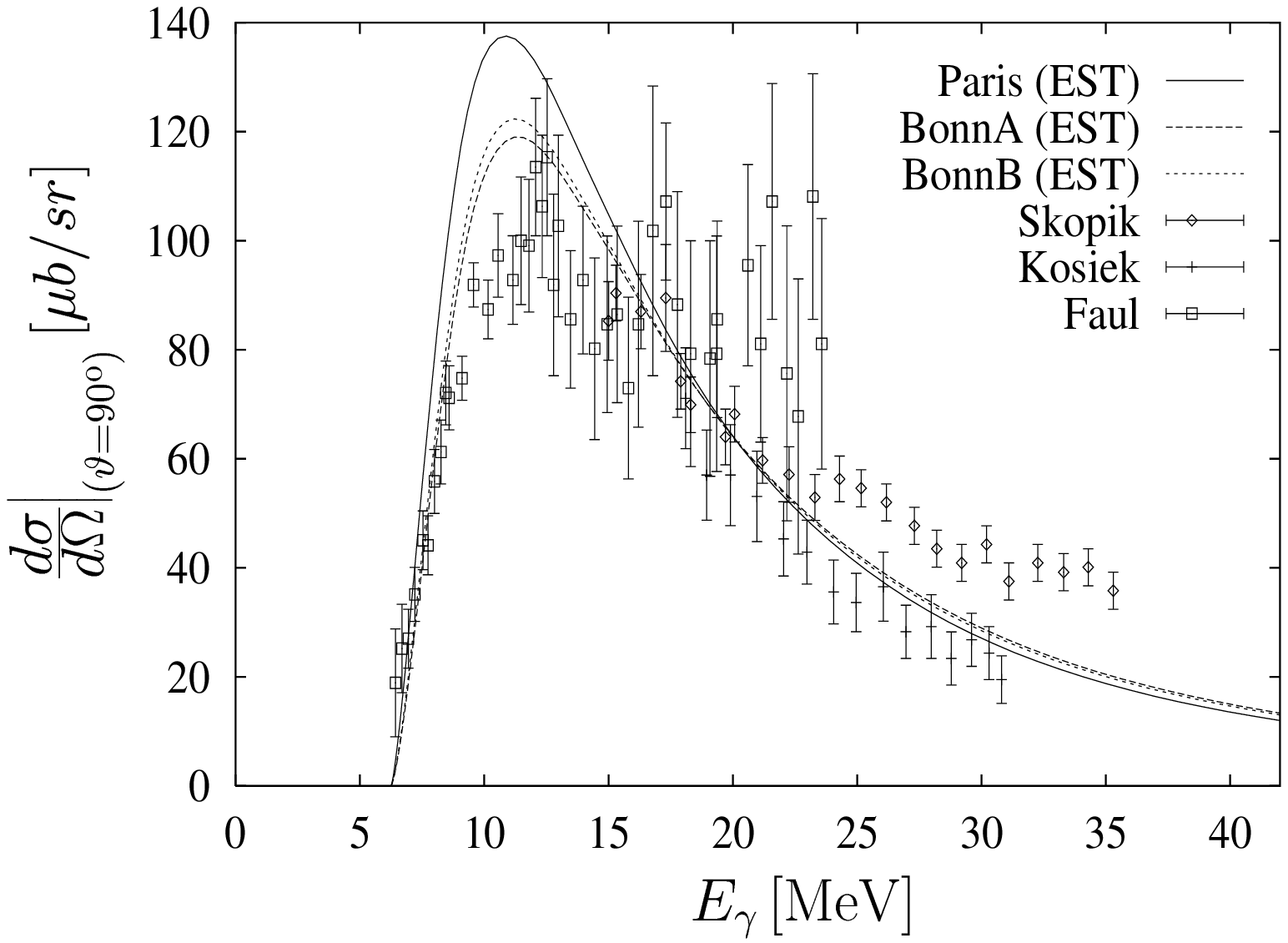,width=49mm} \vspace{-5mm}
\caption{\label{figcross05} Differential cross section at $\vartheta = 90^{0}$
  for $\gamma + \,^{3}$H$\rightarrow n + d$ obtained for the PEST, BAEST and
  BBEST ($j \leq 1^+$) potentials. The data are from
  \protect{\cite{Kosi66,Faul80,Skop81}}.}
\end{minipage} 
\hspace{5mm}
\begin{minipage}[t]{7.5cm}
  \psfig{file=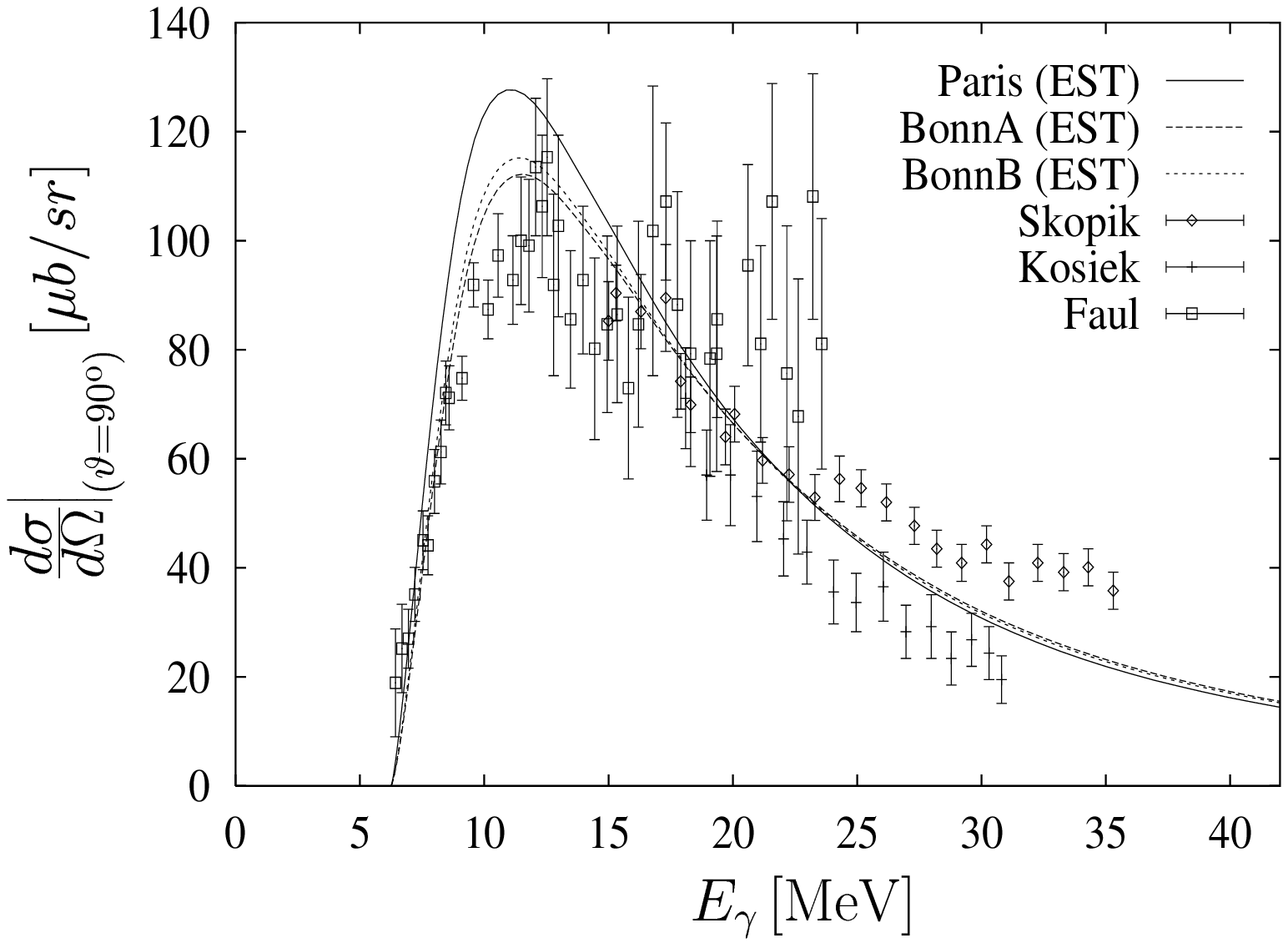,width=49mm} \vspace{-5mm}
\caption{\label{figcross18} Same as Fig. \ref{figcross05} but with
  $(j \leq 2)$ contributions in the potentials.}
\end{minipage}\\ 
\vspace{10mm}\\
\begin{minipage}[t]{7.5cm}  
  \psfig{file=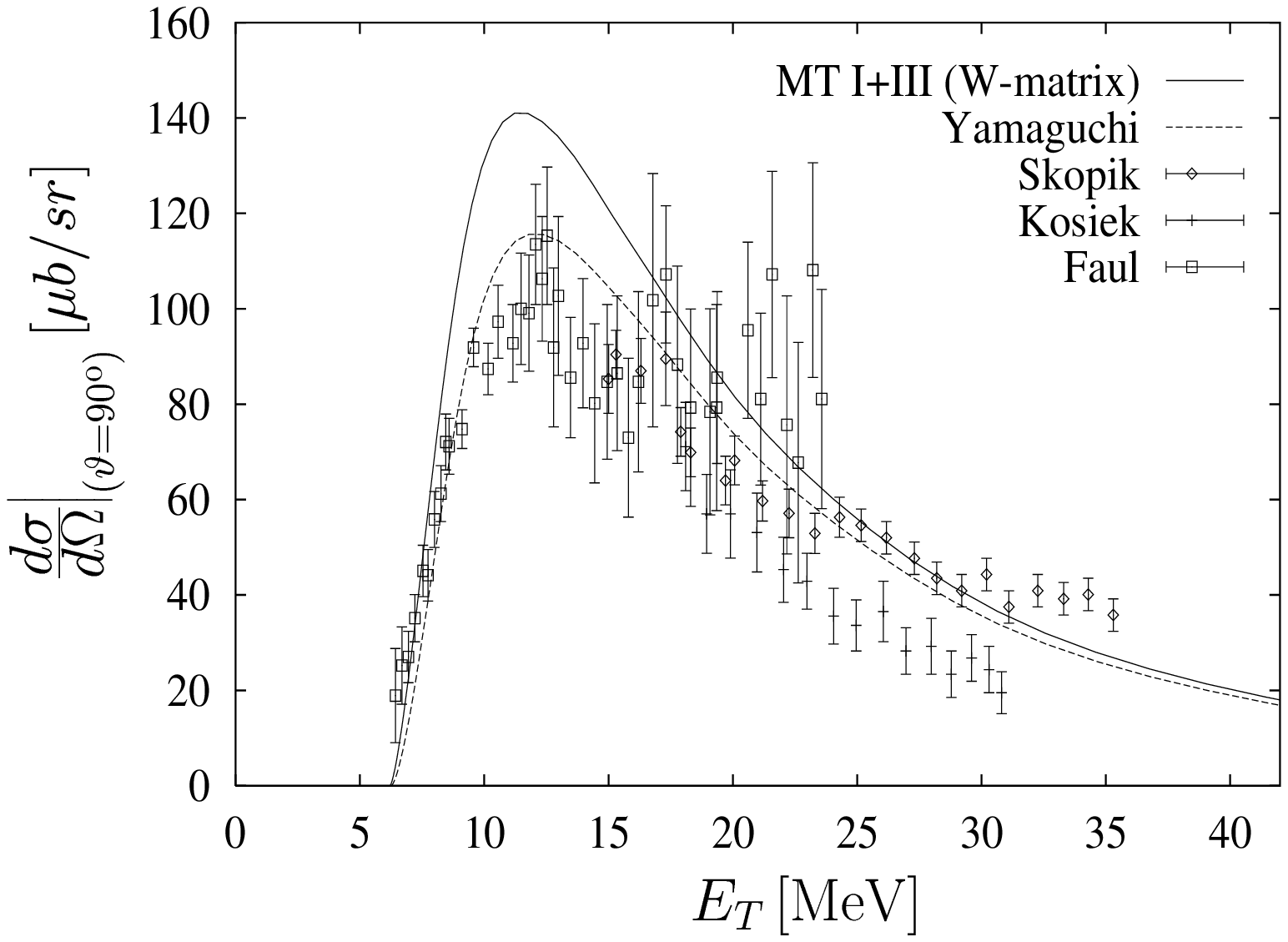,width=49mm} \vspace{-5mm}
\caption{\label{figcrossmtyama} Same as Fig. \ref{figcross05} but for the 
  MT I+III and Yamaguchi potentials.}
\end{minipage} 
\hspace{5mm}
\begin{minipage}[t]{7.5cm}
  \psfig{file=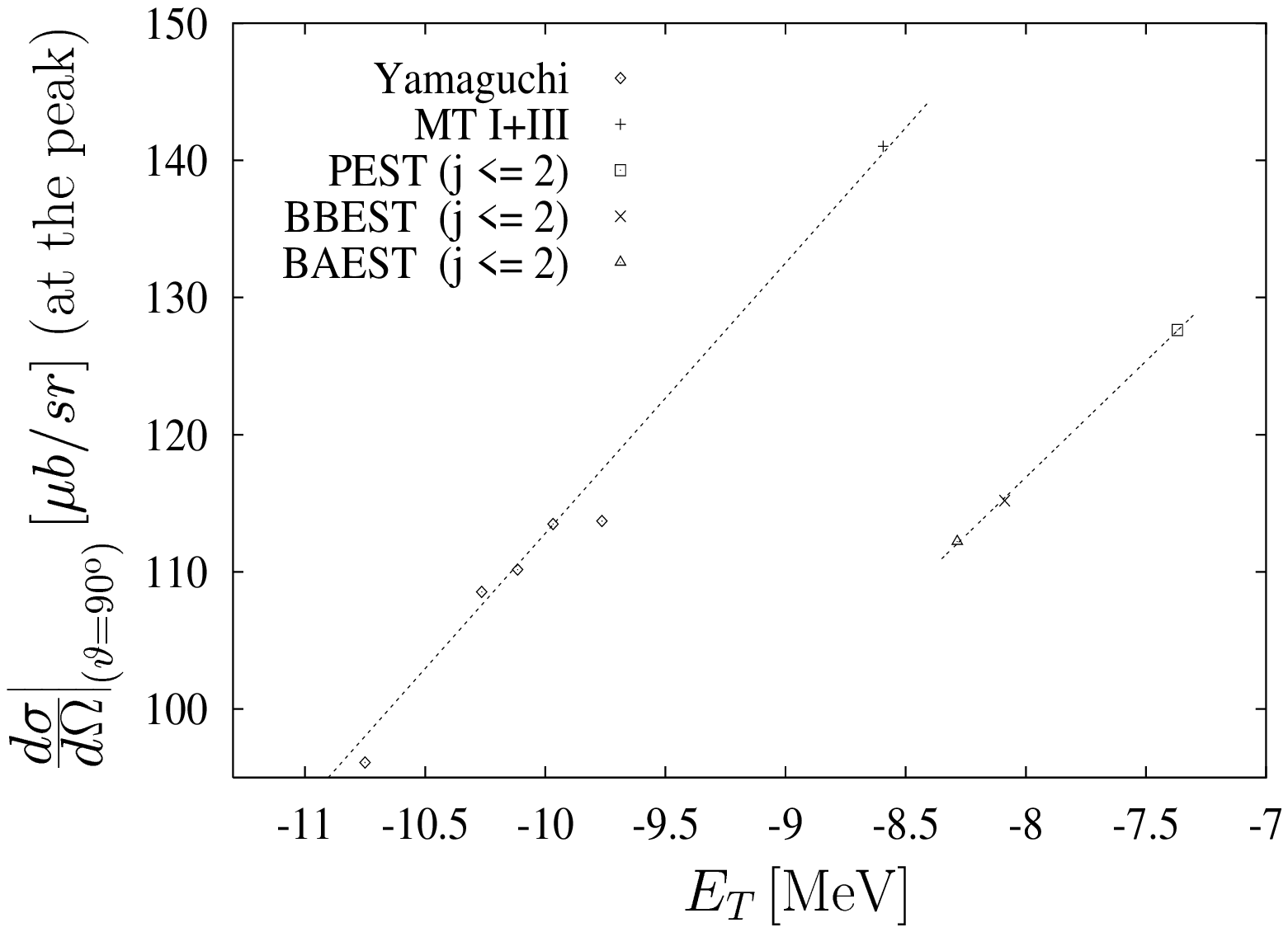,width=49mm} \vspace{-5mm}
\caption{\label{ssline} Correlation of the peak heights and the triton binding
  energies for different potentials.}
\end{minipage}
\end{figure}

According to Fig.~\ref{figpotcomp}, there is a remarkable potential dependence
in the peak region, which appears to open the possibility of testing the
potentials employed. Unfortunately, as seen in Figs.
\ref{figcross05}-\ref{figcrossmtyama}, just in this sensitive region the
experimental errors do not allow for a real judgment. This situation
definitely calls for new, more accurate measurements.

What has to be asked for is, of course, whether this potential dependence is
correlated with other three-body observables. Fig. \ref{ssline} shows the peak
heights of the cross sections versus the corresponding triton binding
energies.  In this plot the results for Yamaguchi potentials with different
parameters and for the Malfliet-Tjon (MT I+III) potential lie on one straight
line, the results for the realistic Paris, Bonn {\sl A} and Bonn {\sl B}
potentials on another, however parallel, line. This strict separation
of the two groups of potentials is not really a statement about "realistic" or
"non-realistic", but is due to the fact that Yamaguchi and Malfliet-Tjon are
$s$-wave 
potentials, while the other ones contain $p$-wave components. In fact,
switching on or off parts of these components, one can shift the realistic
line towards the Yamaguchi-MT line
\cite{Schadowcontribution,Schadowtobepublished}.  In any case, the correlation
exhibited in Fig.  \ref{ssline} is one of the typical features of few-body
physics, as known from the Phillips \cite{Phill68a}or Tjon \cite{Tjon75a}
plots.

At higher energies, i.e., above the peak region, the differences between the
cross sections for the Yamaguchi and MT potentials , and between
the ones for the  realistic potentials vanish. However, there is
a clear discrepancy between both groups of potentials. It should be emphasized
that this discrepancy did not show up in our previous calculations
\cite{Schad97a}, due to the incomplete treatment of relative partial waves
mentioned above. While in these previous calculations all potentials seemed to
favor in this energy region the measurements by Skopik et al. \cite{Skop81}
(this is still true for Yamaguchi and MT), just the realistic ones lie now
between these data and the ones by Kosiek et al. \cite{Kosi66}. This is
another point where a distinct difference between potentials with and without
$p$-wave contributions shows up.  Also in this respect new measurements,
hence, appear desirable.

For completeness it should be mentioned that the role of $p$-wave
contributions in the two-body input was emphasized already by Fonseca and
Lehman \cite{Fons93a}. For an investigation of the dependence on the $d$-state
probability in the deuteron, with Yamaguchi-type interactions in these higher
partial waves, see \cite{Fons92b}.

\subsection{Radiative capture}

\begin{figure}[hbt]
\begin{minipage}[t]{7.5cm}  
  \psfig{file=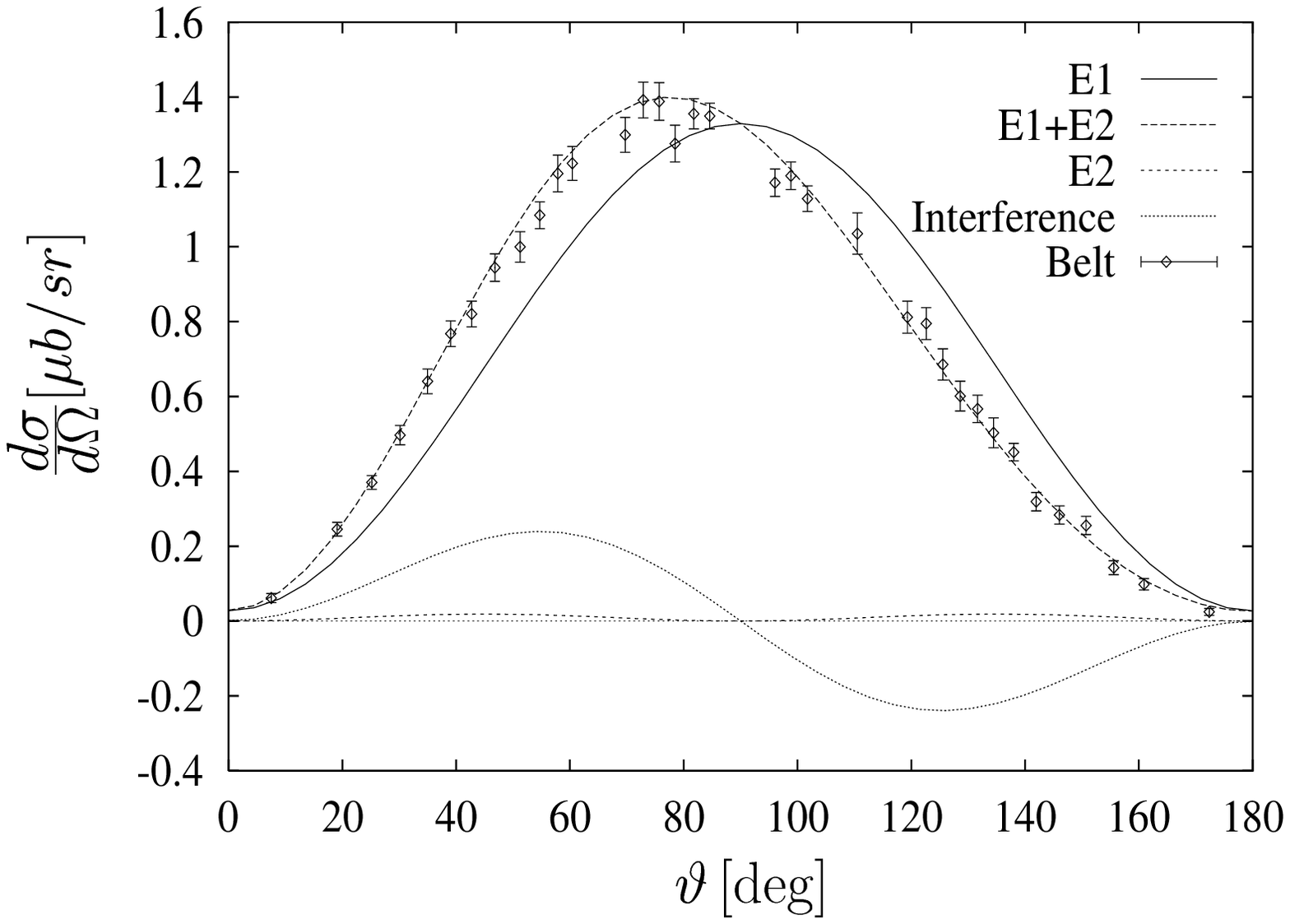,width=49mm} \vspace{-5mm}
\caption{\label{figpdcap1}Cross section for $p$-$d$ capture for the
  PEST ($j \leq 2$) potential at E = 12.1 MeV. The data are from
  \protect{\cite{Belt70}}}
\end{minipage} 
\hspace{5mm}
\begin{minipage}[t]{7.5cm}
  \psfig{file=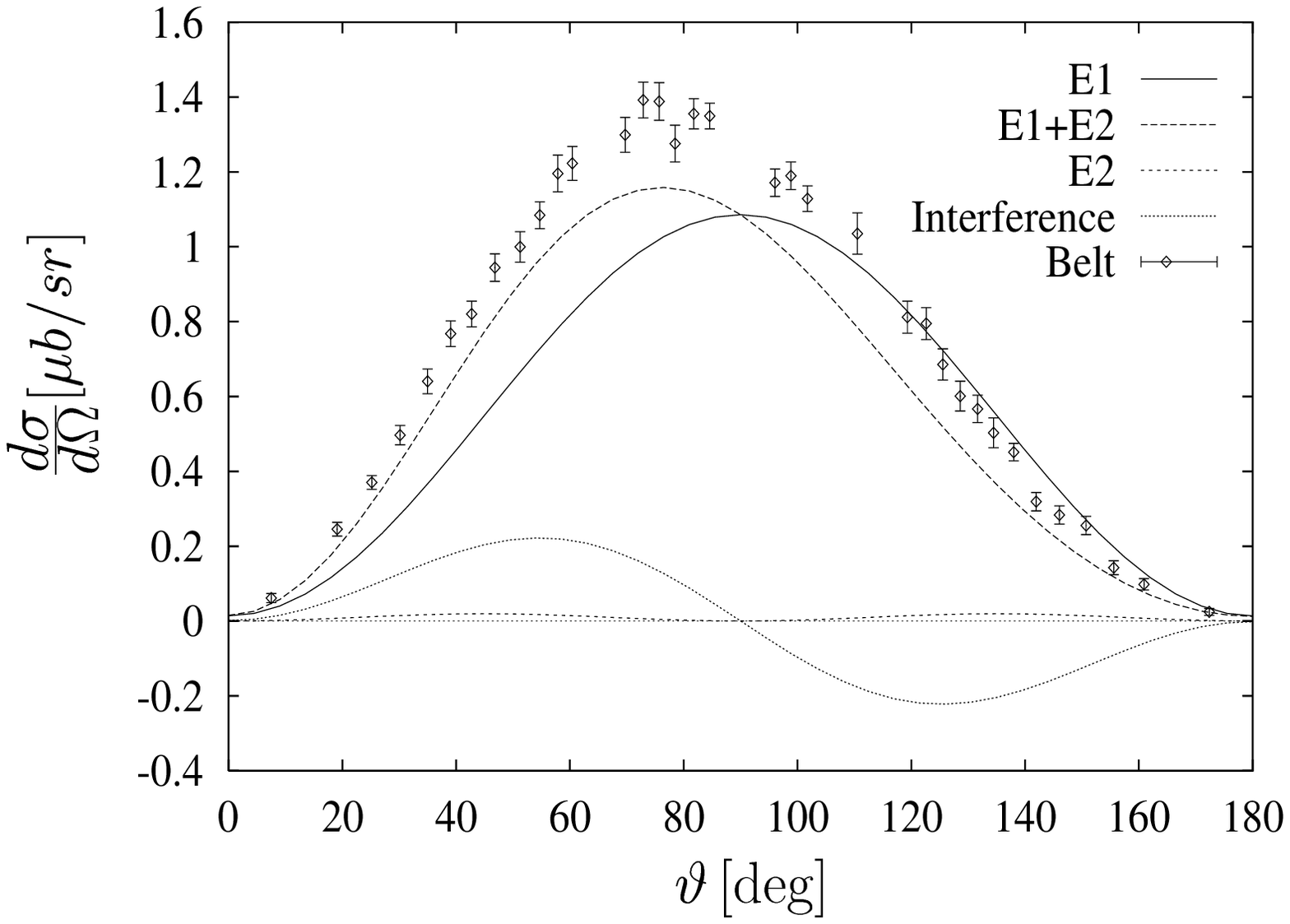,width=49mm}
  \vspace{-5mm}
\caption{\label{figpdcap2} Same as Fig. \ref{figpdcap1} but
  in plane wave (Born) approximation.}
\end{minipage}
\end{figure}

Detailed balance implies that the solutions of (\ref{eq3bphotoint}), i.e., the
amplitudes for the triton or $^3$He photodisintegration into two fragments,
determine also the inverse processes, the $n$-$d$ and $p$-$d$ radiative
capture.  In the above considerations of the total cross section it was
sufficient to take into account only the $E1$ part of $H_{em}$.  Fig.
\ref{figpdcap1} shows, in fact, that the $E2$ contribution, which enters the
total cross section additively, is negligibly small.

This is by no means the case for the interference term of the big $E1$ and the
small $E2$ amplitude occurring in the differential cross section. The $E2$
contribution, hence, is quite relevant, and evidently needed to achieve the
remarkable agreement between theory and experiment. Let us mention that the
situation is quite different in $n$-$d$ capture. There, due to isospin
factors, the interference term is strongly suppressed, and does not lead to
any noticeable correction of the $E1$ result.

Figure \ref{figpdcap2} shows the same $p$-$d$ process, calculated however in
plane-wave (Born) approximation. The general features, in particular the $E2$
effect, are rather similar to the ones of the full calculation.  But, due to
the neglect of FSI, the curves lie clearly below the data. 

\begin{figure}[hbt]
\begin{minipage}[t]{7.5cm}  
  \psfig{file=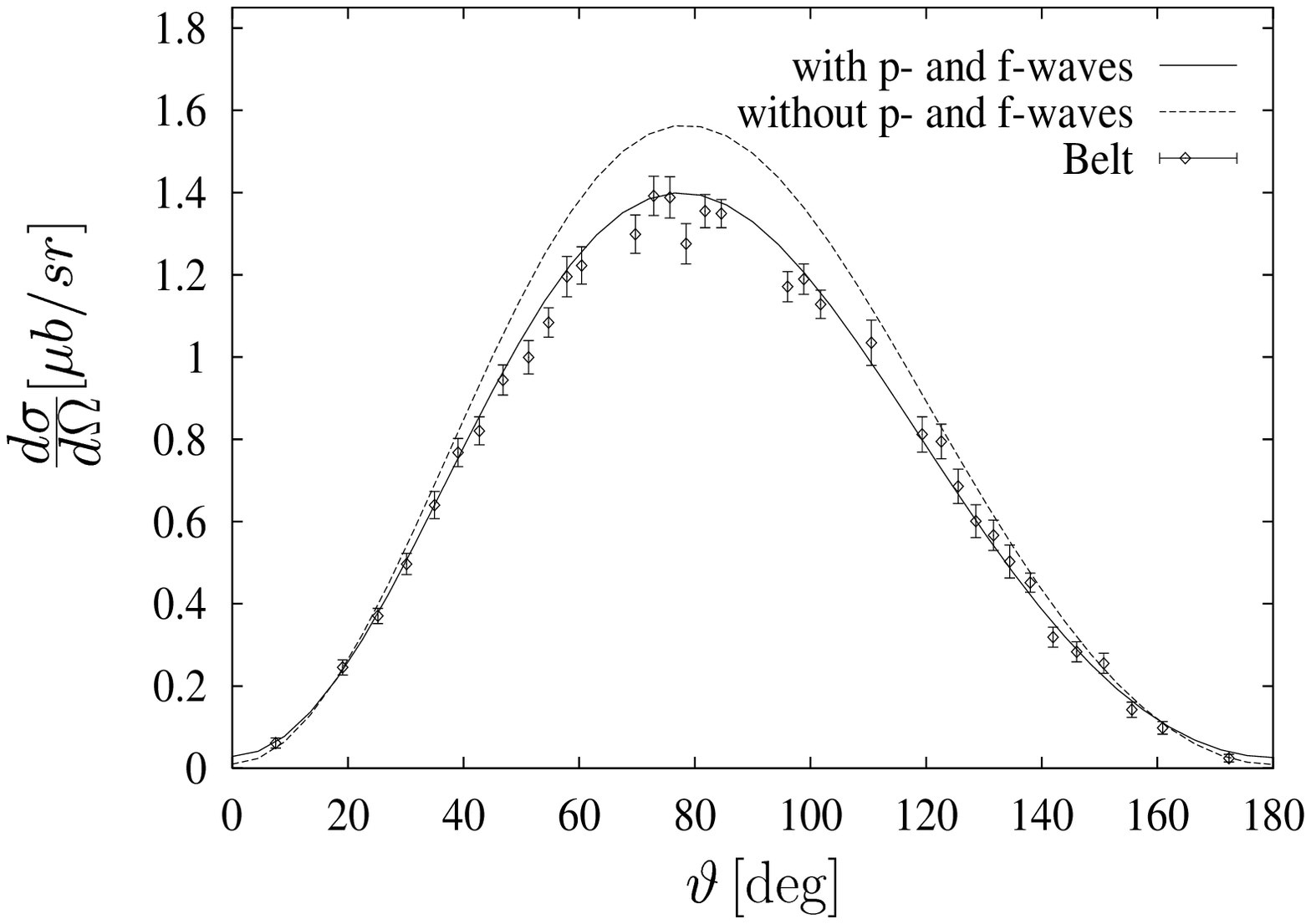,width=49mm} \vspace{-5mm}
\caption{\label{figcappwaves}Differential ross section for $p$-$d$ capture
  with and without $p$- and $f$-wave contributions.}
\end{minipage} 
\hspace{5mm}
\begin{minipage}[t]{7.5cm}
  \psfig{file=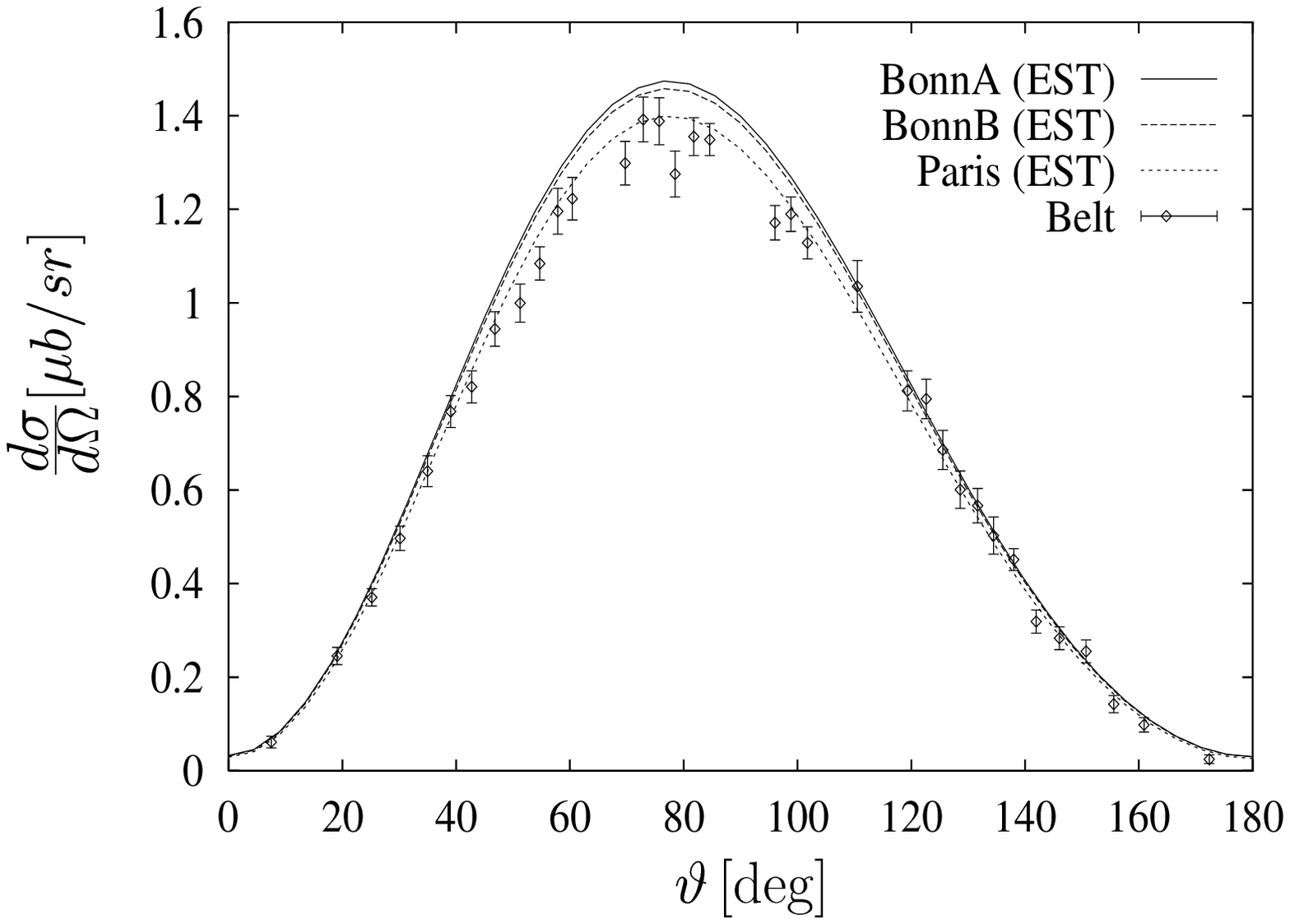,width=49mm}
  \vspace{-5mm}
\caption{\label{figcappotcomp}Full calculation of $p$-$d$ capture at
  E = 12.1 MeV for the PEST, BAEST, and BBEST potentials($j \leq 2$).}
\end{minipage}
\end{figure}

In Fig. \ref{figcappwaves} it is shown that agreement between theory and
experiment is only achieved when incorporating the $p$- and $f$-wave
contributions.  Their relevance has been pointed out already in the
photodisintegration case.  To compare the corresponding observations, we have
to bear in mind that the energy of 12.1 MeV corresponds to 
photodisintegration above the peak region. There, we also found a lowering of
the cross sections for potentials with $p$-wave contributions which, however,
led to results between the sets of data.  The excellent agreement between
theory and experiment in the present case of radiative capture points at the
high accuracy of the 
experimental data, as well as  the reliability of the calculations.

In the photodisintegration there was practically no potential dependence at
somewhat higher energies, and thus, as expected, the difference between the
Bonn {\sl A}, Bonn {\sl B} and Paris results shown in Fig. \ref{figcappotcomp}
is also rather small.

In our review we tried to give a consistent picture of the influence of FSI,
MEC, the potential dependence, the effect of $p$-waves in the interaction, or
the $E2$ contribution in $H_{em}$, on the cross sections of three-nucleon
photodisintegration or radiative capture. It should be mentioned that
such questions have been asked also with respect to polarization observables
in radiative capture \cite{Fons93a,Fons92b,Fons91a,Ishi92b}. In the
calculations by Fonseca and Lehman, e.g., a strong FSI and $p$-wave effect was
found for the tensor analyzing power $T_{20}$ and $A_{yy}$.

Very low energy radiative capture processes are of considerable astrophysical
relevance. The $n$-$d$ radiative capture, which at such energies is almost
entirely a magnetic dipole ($M1$) transition, was studied in plane wave (Born)
approximation by Friar et~al. \cite{Friar90}. In these investigations the
authors employed their 
configuration-space Faddeev calculations of the triton wave function, with
inclusion of three-body forces and pion exchange currents.  Various trends,
e.g. the correlation between cross sections and triton binding energies, and
its potential dependence were pointed out.

More recently a rather detailed investigation of such processes has been
performed by Viviani et~al. \cite{Viviani96a}. Their calculations  employed
the very accurate 
three-nucleon bound- and continuum states, obtained in the
variational pair-correlated hyperspherical method developed, tested, and
applied over years by this group. For a review we refer to the invited talk
by Viviani at this conference \cite{Viviani97a}.

\section{FOUR-BODY PROCESSES}

\setcounter{equation}{0}

In this section we recall the generalization of the three-body momentum space
approach to the four-body case, and review its applications to the
photoprocesses $\gamma + \, ^4$He $ \rightarrow n + \,^3$He or $p + \,^3$He.
Similarly to the three-body case, we study the influence of FSI, MEC, the
potential dependence, or $E2$ contributions on the corresponding total and
differential cross sections.

\subsection{Formalism}

The generalization of the AGS equations (\ref{eqags}) is the set of $18 \times
18$ operator identities \cite{Grassberger67a}

\begin{equation}
U^{\sigma \rho}_{\beta \alpha} = (1 - \delta_{ \sigma \rho}) \,
\delta_{\beta \alpha} \, G^{-1}_0 \,
T^{-1}_{\alpha} \, G^{-1}_0 + \sum\limits_{\tau \neq \sigma}
\sum\limits_{\gamma} U^{\tau}_{\beta \gamma} \,G_0\, T_{\gamma} \,G_0\,
U^{\tau \rho}_{\gamma \alpha}
\end{equation}

\noindent
for the generalized four-body transition operators $U^{\sigma \rho}_{\beta
  \alpha}$. As in the AGS equations, this set contains in its kernel the
two-body operators $T_{\gamma} = T_{i j}$. But now there occur also the
three-body operators $U^{\tau}_{\beta \alpha}$ given by (\ref{eqags}). The
indices $\rho, \sigma, \tau$ denote the three-body subsystems in $(3+1)$
fragmentations or the possible $(2 +2)$ fragmentations. As an equation for
four-body quantities it acts on three relative (Jacobi) variables being thus
three-dimensional after partial wave decomposition.

Inserting the separable expansion (\ref{eqtmatsep}), its dimension is reduced,
as usual, by one. That is, we now end up with an effective two-body equation
of exactly the AGS form (\ref{eqags}),

\begin{equation}
{\cal U}^{\sigma \rho} = (1 - \delta_{\sigma \rho}) {\cal G}^{-1}_0
+ \sum\limits_{\tau \neq \sigma} {\cal T}^{\tau}\, {\cal G}_0 \,
{\cal U}^{\tau \rho} \,.
\end{equation}

\noindent
As in the genuine three-body case it may be, and has been used as it stands,
i.e., as a two-dimensional relation. But again it appears more convenient to
further reduce the dimension by expanding also the operator ${\cal T}^{\tau}$
into separable terms.  This two-step reduction scheme, concerning both the
two- and three-body subsystem operators in the four-body equations, provides
us then with an effective two-body matrix equation

\begin{equation}
\label{eq4btmatint}
{\cal T} ({\vec q}^{\, \prime},{\vec q}\,) = {\cal V} ({\vec q}^{\, \prime},
 {\vec q}\,) + \int \! d^{3}{ q^{\, \prime \prime}} \, {\cal V} 
({\vec q}^{\, \prime \prime}) \, {\cal G}_0 (q^{\, \prime \prime}) \,
{\cal T}  ({\vec q}^{\, \prime \prime}, {\vec q}\,) 
\end{equation}

\noindent
of the LS form (\ref{eqtmatint}) or (\ref{eq3btmatint}), of course with a much
more complicated structure of the potentials ${\cal V}$ and Green functions
${\cal G}_0$. Here, ${\vec q}$ and ${\vec q}^{\, \prime}$ denote the relative
momenta between the clusters in the $(3+1)$ or $(2+2)$ channels.

And, analogously to the steps leading in the two- and three-body cases to the
photodisintegration equations (\ref{eqphotoint}) and (\ref{eq3bphotoint}), we
end up with their four-body counterpart \cite{Casel80a}

\begin{equation}
{\cal M}({\vec q}\,) = {\cal B} ({\vec q}\,) + \int \! d^3 q^{\, \prime} \,
{\cal V}
({\vec q}, {\vec q}^{\, \prime}) \, {\cal G}_0 ({\vec q}^{\, \prime}) \,
{\cal M} ({\vec q}^{\, \prime},{\vec q}\,)
\end{equation}

\noindent
for an off-shell extension of the (properly antisymmetrized) four-body
photodisintegration amplitude

\begin{equation}
M ({\vec q}\,) = 2 \;{^{(-)}} {\langle} {\vec q}; \psi_{III} | H_{em} | 
\psi_{IV} \rangle \,.
\end{equation}

\noindent
Again the effective potentials and Green functions can be taken over from the
purely nuclear case, i.e., from Eq. (2.3), and  the inhomogeneity 
${\cal B} ({\vec  q}\,)$ is an off-shell extension of the plane-wave
 (Born) approximation of this process,

\begin{equation}
B ({\vec q}\,) = 2 \, {\langle} {\vec q \,}| \langle \psi_{III} | H_{em} | 
\psi_{IV} \rangle \,.
\end{equation}

In what follows we show results obtained for the Yamaguchi potential and the
semi\-realistic Malfliet-Tjon (MT I+III) potential. For the treatment of the
two-and three-body subsystem operators in the kernel of the four-body
equations, the $W$-matrix approximation \cite{Bart86} and the energy dependent
pole expansion (EDPE) \cite{Sofianos79a} were chosen. In the pure nuclear case
these two approximations have led to rather accurate results. The high
accuracy of the $W$-matrix representation in the photonuclear case, moreover,
was demonstrated in Sec.~1.4 (see, in particular, Fig. 4).  The accuracy of
the EDPE is, of course, less evident and involves, in fact, one of the
uncertainties of the present four-nucleon momentum-space approach.

The electromagnetic $E1$ and $E2$ operators, with and without Siegert
replacement, are the ones given in Eqs. (\ref{eqhem1}), (\ref{eq3bmmec}),
(\ref{eqhem2}) and (\ref{eqhem2mec}).  The sums runs now, of course, till
$j=4$.

\subsection{Results}

\begin{figure}[hbt]
\begin{minipage}[t]{7.5cm}  
  \vspace{-7mm} \psfig{file=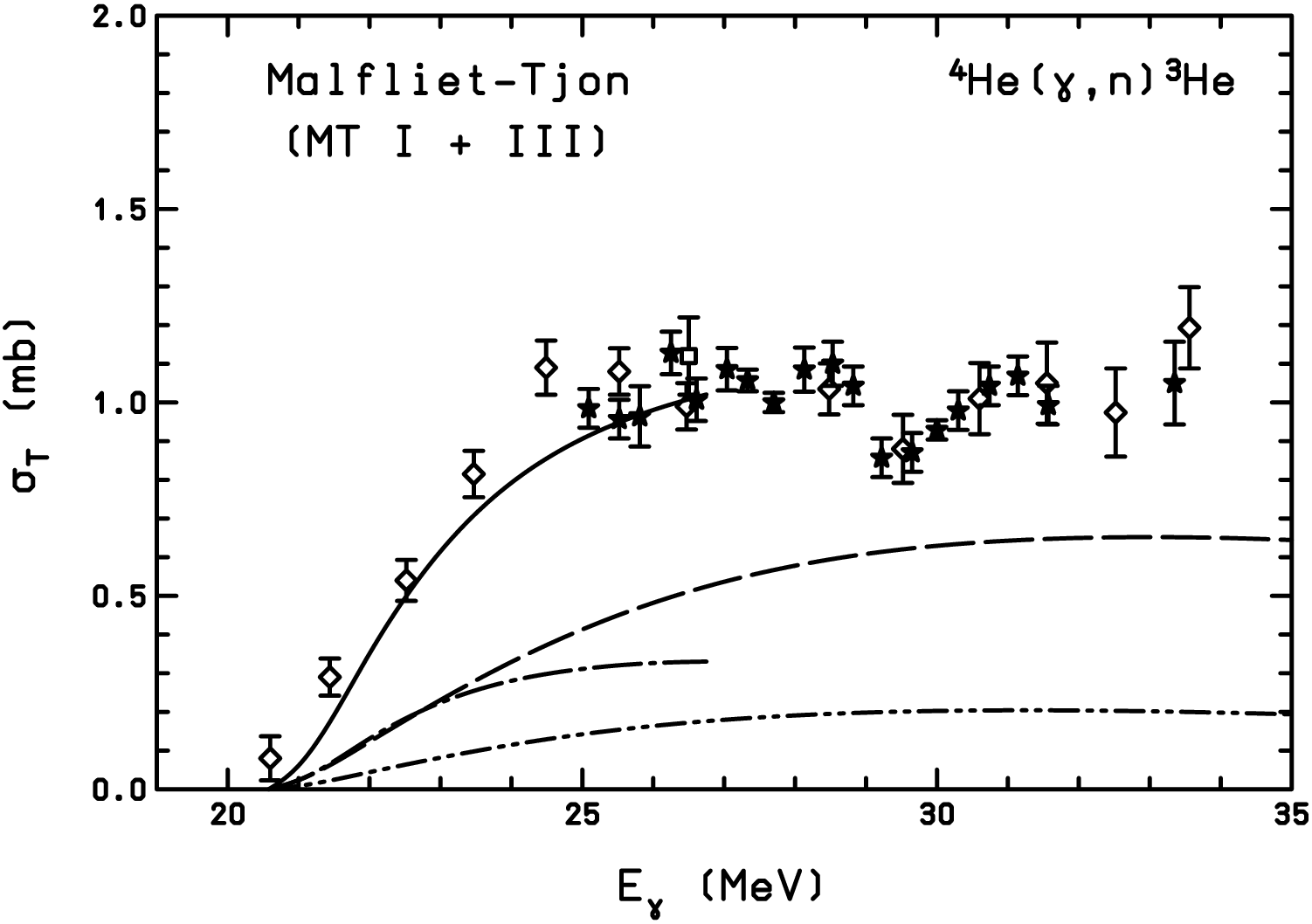,width=74mm} \vspace{-9mm}
\caption{\label{gamfig1} Total cross section for 
  $^4$He$(\gamma,n)\,^3$He at 
  low energies: Full solution with exchange currents (------) and without
  ($-\cdot-$); plane wave approximation with exchange currents ($--$) and
  without ($-\cdot\cdot-$); the data are from \cite{Berman80a} $\diamond$,
  \cite{Ward81a} $\star$, and \cite{Asai94a} $\Box$.}
\end{minipage} 
\hspace{5mm}
\begin{minipage}[t]{7.5cm}
  \vspace{-7mm} \psfig{file=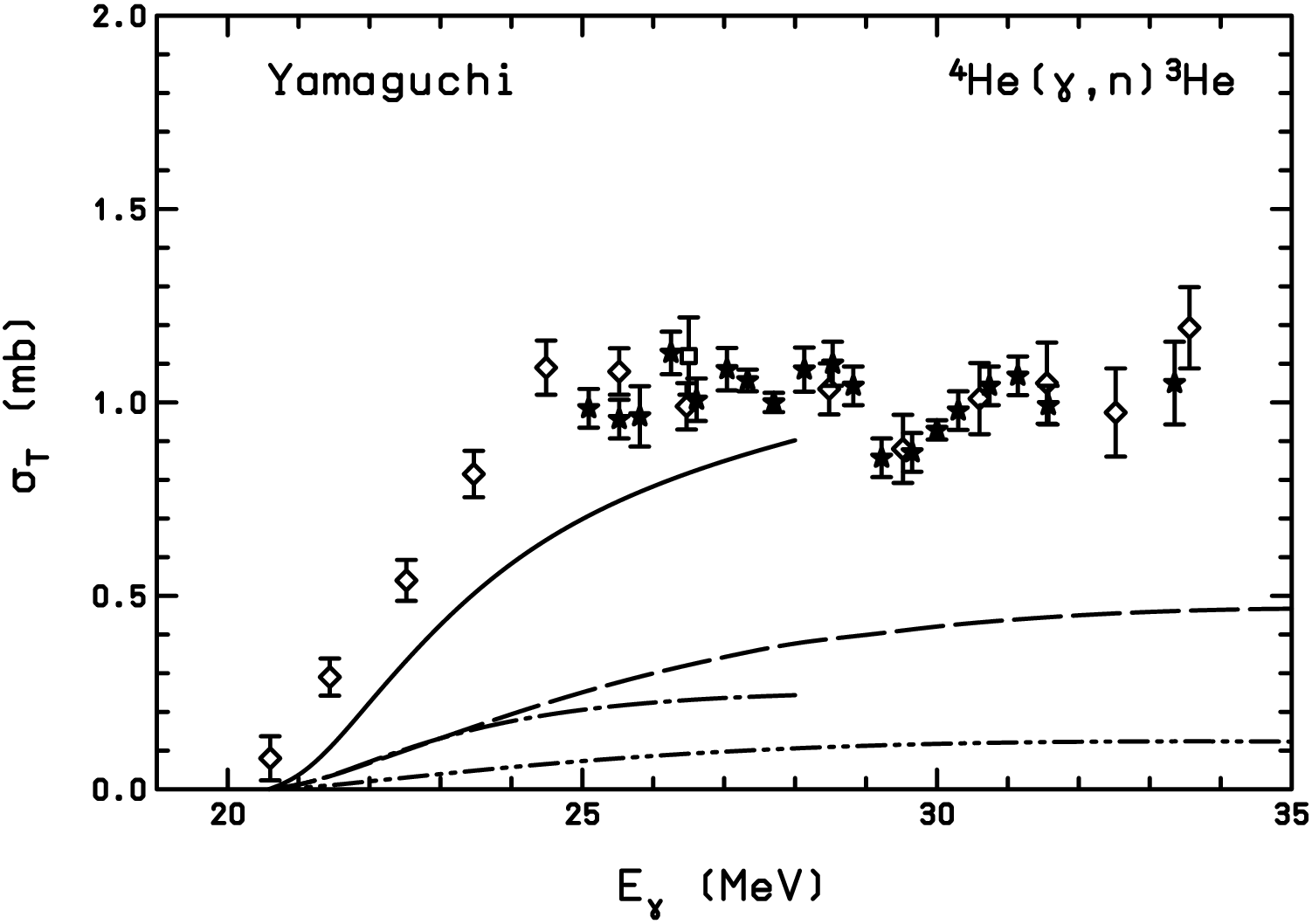,width=74mm} \vspace{-9mm}
\caption{\label{gamfig2}Same as Fig. \ref{gamfig1}, but with the Yamaguchi
  potential.}
\end{minipage}
\vspace{-3mm}
\end{figure}

Figure \ref{gamfig1} shows the total $^4$He$(\gamma,n)\,^{3}$He cross sections
for the MT potential. Due to the complicated cut structure of the integral
equation kernel, the full calculations could be performed only below the
three-fragment break-up threshold. Even more pronounced than in the three-body
case (compare Fig. \ref{figtritonmec}) is the difference between the full and
the Born, the MEC and the non-MEC results. Note that only with inclusion of
these effects the remarkable agreement between theory and experiment is
achieved \cite{Ellerkmann95b}.

Fig. \ref{gamfig2} shows the corresponding results for the Yamaguchi
potential. They lie below the MT curves, in fact too low as compared to the
data. This potential dependence resembles completely the one exhibited in the
three-body case.  Again, a stronger $^4$He binding (39.1 MeV for Yamaguchi as
compared to 30.1 MeV for MT I+II) leads to a lowering of the cross section in
the peak region.

\begin{figure}[hbt]
\begin{minipage}[t]{7.5cm}  
  \vspace{-7mm} \psfig{file=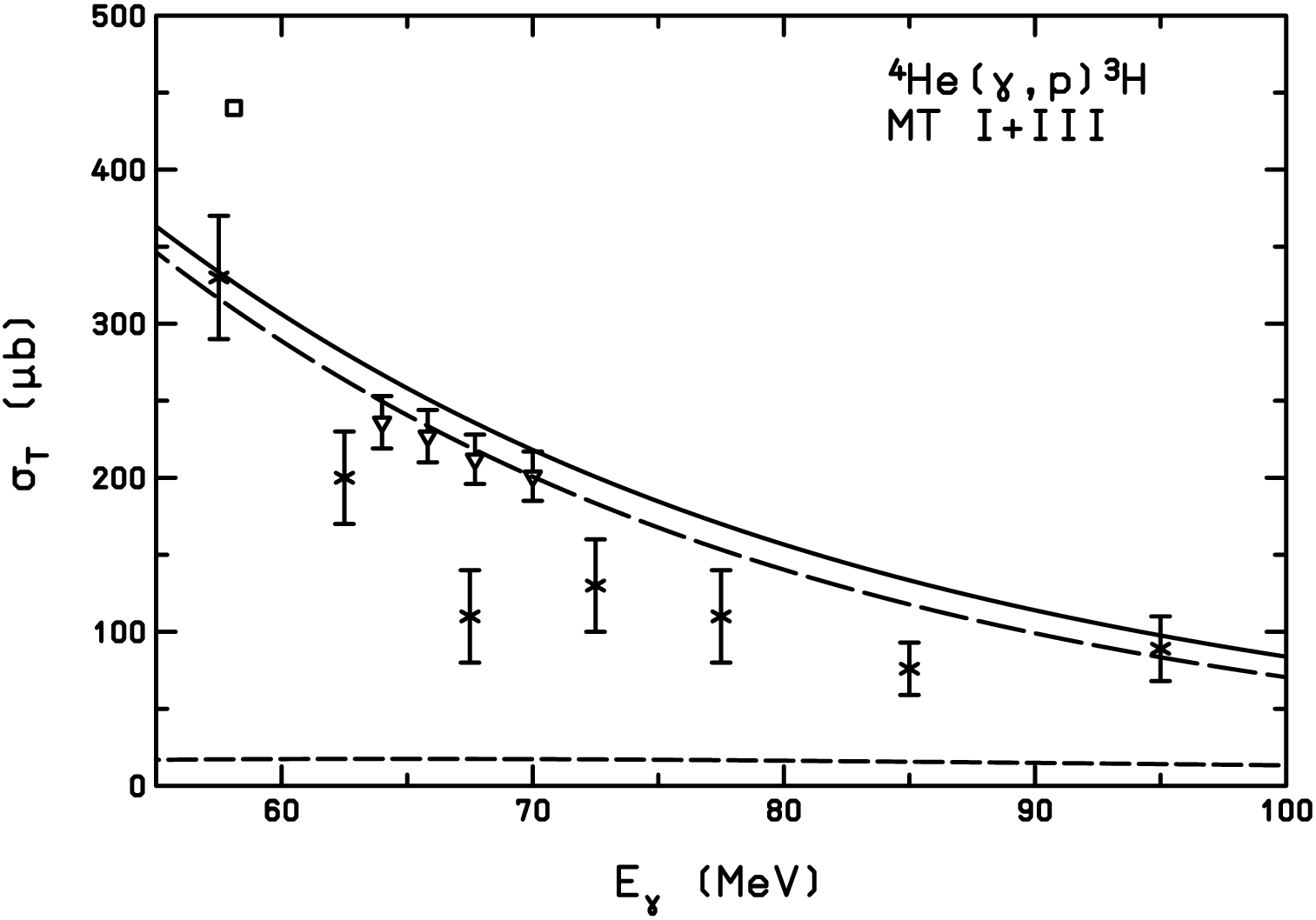,width=74mm} \vspace{-9mm}
\caption{\label{gamfig3} Total cross section for $^4$He$(\gamma,p)\,^3$H
  at higher energies: Plane wave approximation including E1 and E2 (------).
  Contributions of E1 (--- ---) and E2 ($---$); the data are from
  \cite{Jones91a} $\bigtriangledown$, \cite{Gorbunov76a} $\times$, and
  \cite{Bernabei88a} $\Box$.}
\end{minipage} 
\hspace{5mm}
\begin{minipage}[t]{7.5cm}
  \vspace{-7mm} \psfig{file=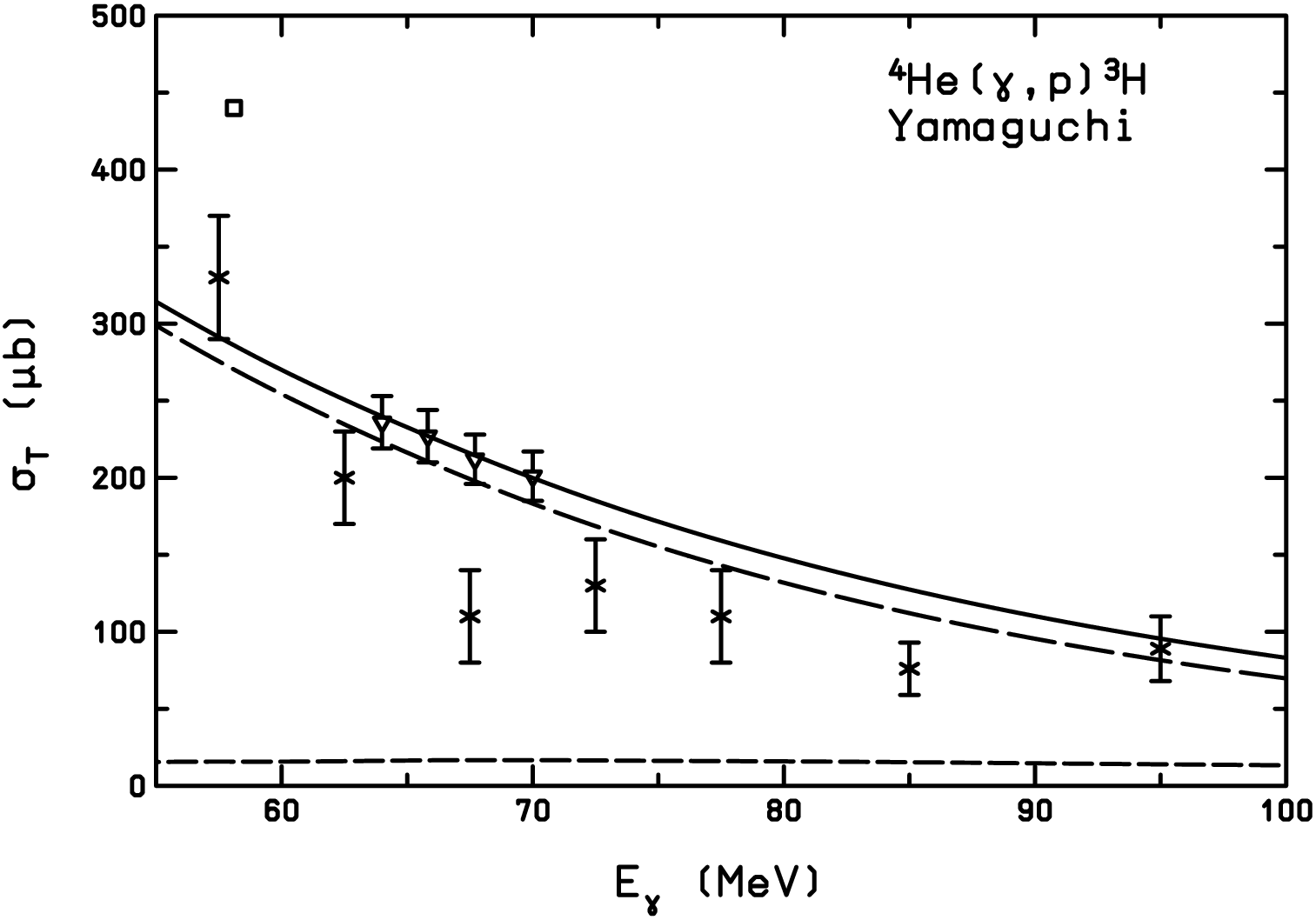,width=74mm} \vspace{-9mm}
\caption{\label{gamfig4}Same as Fig. \ref{gamfig3}, but with the Yamaguchi
  potential.}
\end{minipage} 
\vspace{-7mm}
%\end{figure}
%
%\begin{figure}[hbt]
\begin{minipage}[t]{7.5cm}  
  \psfig{file=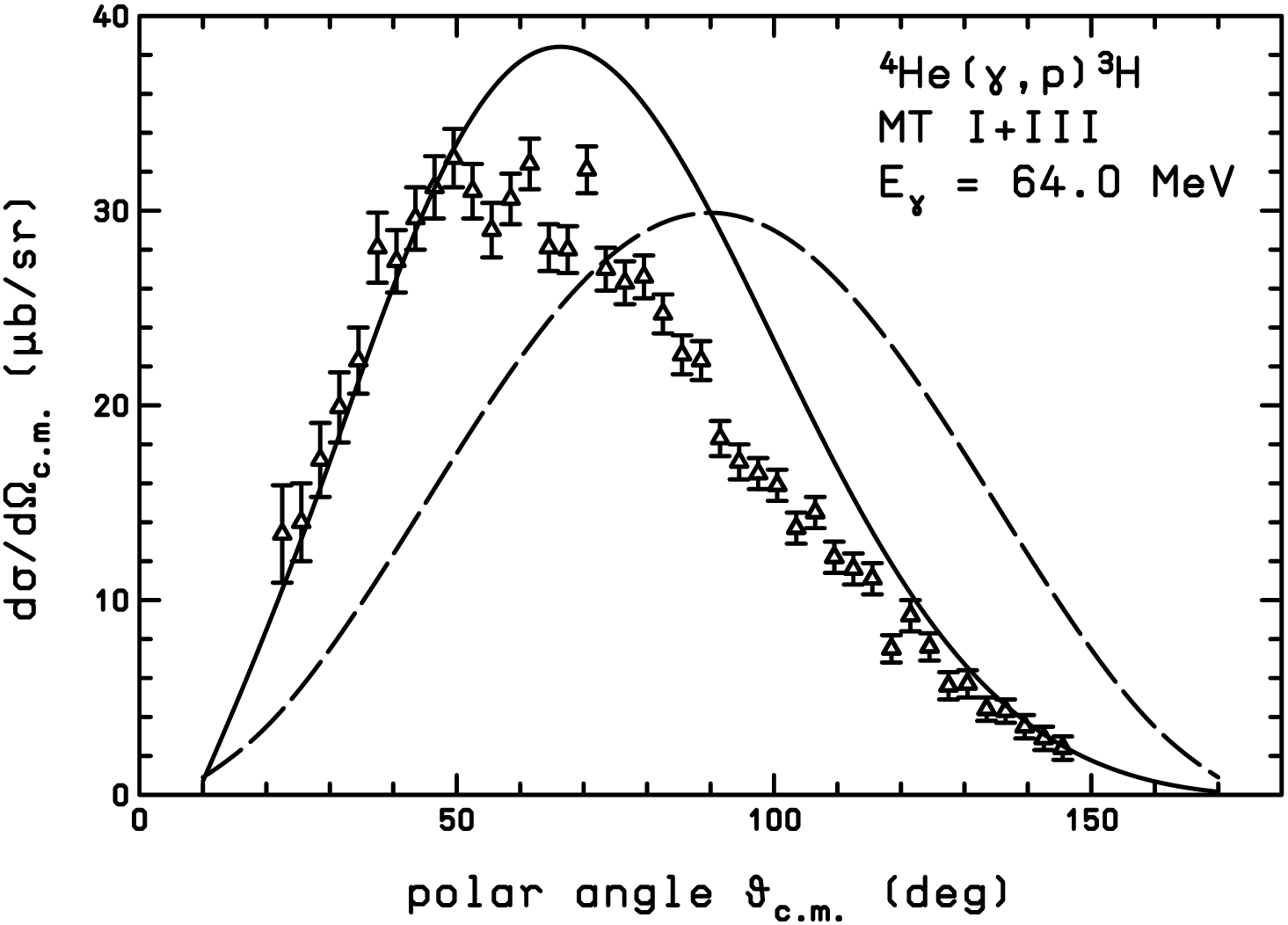,width=74mm} \vspace{-9mm}
\caption{\label{gamfig5}Differential cross section: Plane wave approximation 
  including E1 and E2 (------). Contribution of E1 only ($--$); the data are
  from \cite{Jones91a}.}
\end{minipage} 
\hspace{5mm}
\begin{minipage}[t]{7.5cm}
  \psfig{file=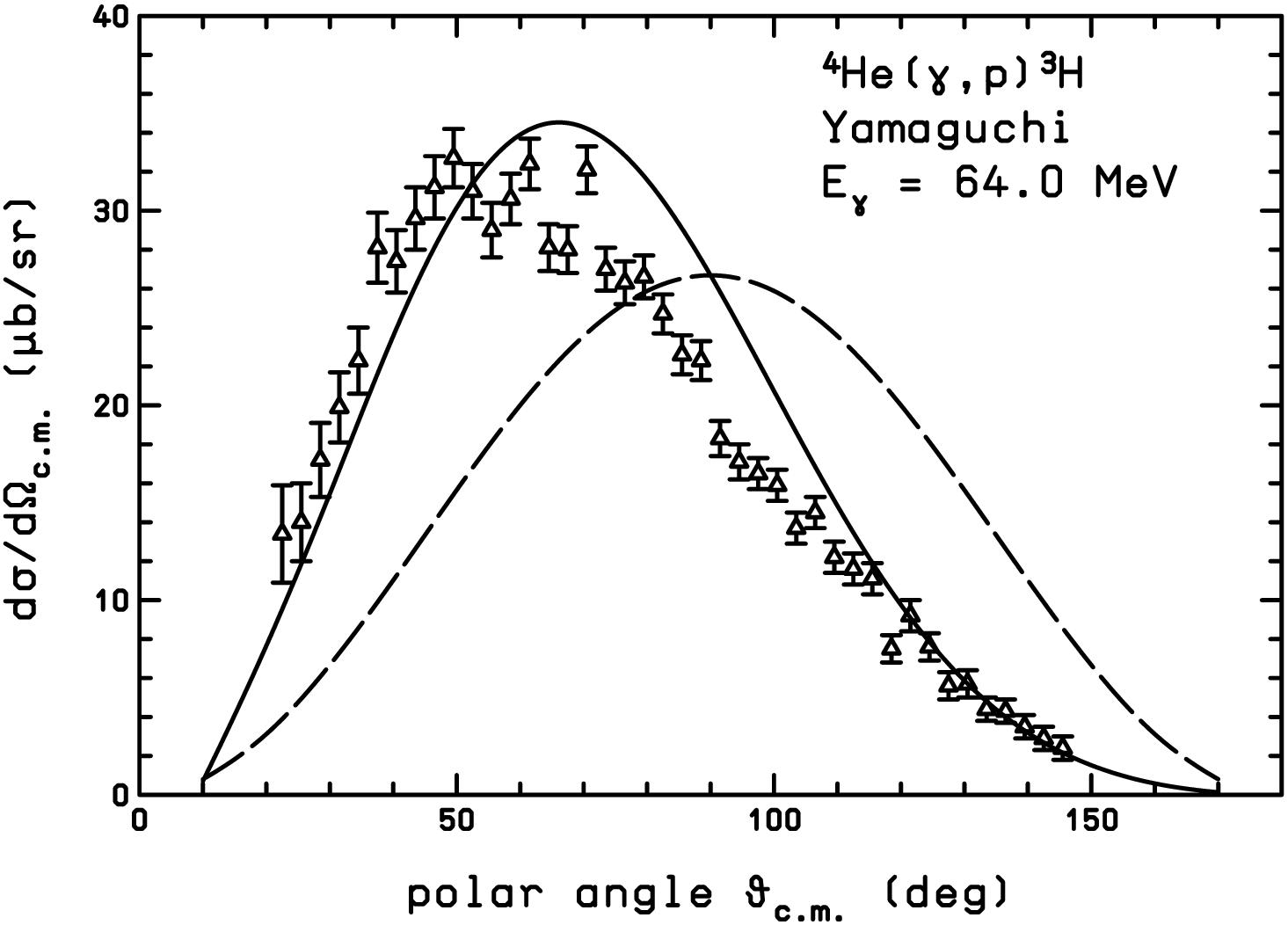,width=74mm} \vspace{-9mm}
\caption{\label{gamfig6}Same as Fig. \ref{gamfig5}, but with the Yamaguchi
  potential.}
\end{minipage}
\end{figure}

According to the three-body experience, FSI contributions can be neglected at
higher energies. There, as shown in Figs. \ref{gamfig3} and \ref{gamfig4},
also the potential dependence vanishes, and the $E2$ contributions are small.
All these observations are supported by the remarkable agreement with newer
experimental data \cite{Jones91a}.  The two- and three-body experiences, on
the other hand, indicate a strong $E2$ effect in the differential cross
sections. And, in fact, the $^4$He$(\gamma,p)\,^{3}$H differential cross
sections in Figs. \ref{gamfig5} and \ref{gamfig6} show $E2$ corrections
similarly to the ones of Figs \ref{figdeut2}, \ref{figpdcap1}, and
\ref{figpdcap2}.  In contrast to the three-body case, there remain, however,
discrepancies which, according to the experiences gained in this case, are
probably to be attributed to the neglect of higher partial waves in the two-
and three-body subsystems. An improvement of the EDPE, the incorporation of
further multipoles or FSI corrections may also be considered in this context.

\subsection{Comments and alternative approaches}

The photodisintegration of $^{4}$He into $n + \,^{3}$He or $p + \, ^{3}$H has
for a long time been a rather controversial topic.  Early data appeared
consistent with the picture of a giant dipole resonance, a feature also
supported by shell model \cite{Halderson81}, and resonating group (RGM)
calculations \cite{Wachter88a} In contrast, applications of the momentum-space
integral equation approach indicated from the very beginning a fairly flat,
non-resonant behavior \cite{Casel80a,Casel86a}, and at the same time a similar
trend was observed also experimentally \cite{Berman80a}.

The drastic separable approximations of the three-body transition operators in
the kernel of the four-body equations, employed in these calculations, have
been replaced in the meantime by the more reliable EDPE, and additional new
measurements are also available.  As seen in Fig. \ref{gamfig1}, all these
developments have confirmed the flat low-energy behavior of the total cross
section predicted in \cite{Casel80a} and \cite{Berman80a}.

As mentioned already, the RGM calculations by Wachter et~al.~\cite{Wachter88a}
reproduced the apparently non-existing giant resonance.  This is not
surprising. Their application of the RGM approach is characterized by a
completely unrealistic potential chosen for purely technical reasons. The
corresponding binding energies, e.g., differ from the experimental ones by
almost a factor of two.  To cure this shortcoming, the authors replaced the
theoretical by the experimental values in the Siegert factor. This
inconsistency allowed them to fit first the giant resonance behavior and then,
by going back to the unrealistic theoretical values, also the new flat results
\cite{Unkel92}. No need to emphasize that this high ambiguity reduces the
predictive power of such a method.

The complexity of the full four-body integral equations suggests to look
for somewhat less ambitious, but none the less
sufficiently reliable models.  The approach suggested in \cite{Sofia93a} is
based on writing the full outgoing scattering state as a product

\begin{equation}
^{(-)}\langle \vec q^{\, \prime}; \psi_{III}|
\simeq \, {^{(-)}}\langle \vec q^{\, \prime}| \langle \psi_{III}| \,.
\end{equation}

\noindent
Here, the two-fragment scattering state $^{(-)}\langle \vec q \, |$ describes
the motion of the outgoing nucleon under the influence of an optical
potential, adjusted to the $n$-$^{3}$He or $p$-$^{3}$H scattering data. A
further simplification consists in calculating the three- and four-body bound
states within the integrodifferential approach (IDEA) \cite{Ripelle88}.  The
results obtained in this 
way for the total cross section are in good agreement with those of the full
treatment.  This model, moreover, allows one immediately go beyond the
three-body break-up threshold (compare the comments on Fig. \ref{gamfig1}).
The Coulomb effect, moreover, is easily incorporated.

Note that the determination of the optical potential between the outgoing
fragments was based in \cite{Sofia93a} on a simple analytical expression
adjusted to the data.  Instead, the more sophisticated Marchenko inversion
technique could be employed, as discussed with respect to
electrodisintegration in Sec.~3.3.

\section{THREE- AND FOUR-NUCLEON ELECTRODISINTEGRATION}

Our main topic is the photodisintegration or radiative capture of light
nuclei. The considerable experimental and theoretical interest devoted to the
corresponding electrodisintegration processes, however, suggests to briefly
review also methods developed, and results achieved in this field, and to
compare them with a recent alternative access to this problem based on
Marchenko inversion.

\subsection{Three-nucleon problem}

\begin{figure}[hbt]
\begin{minipage}[t]{7.5cm}  
  \epsfig{file=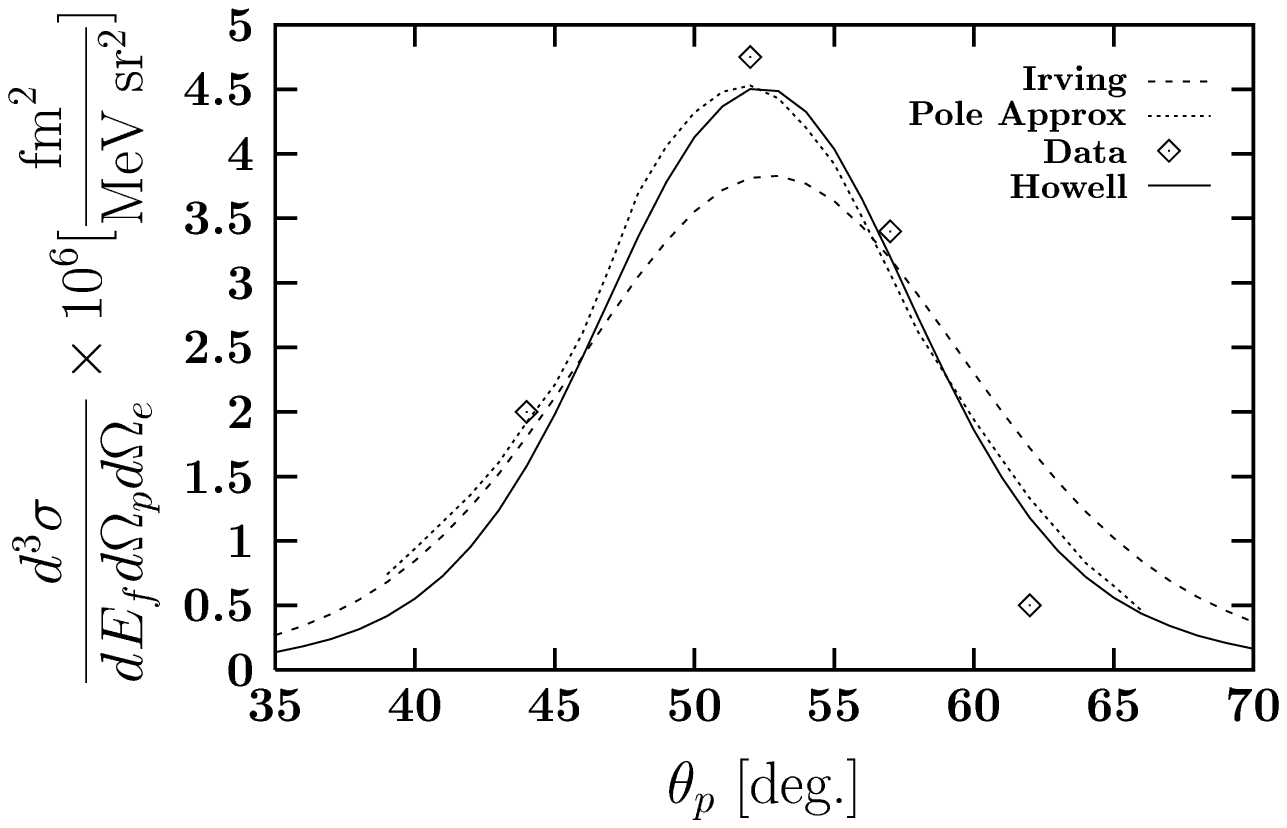,width=7.1cm,height=5.2cm} \vspace{-9mm}
\caption{\label{figelectro1} Coincidence cross section 
  for $e+^3$He$\rightarrow d + p + e'$ as a function of the proton scattering
  angle $\theta_p$ for the kinematics of \cite{Griff64a}. The reanalyzed data
  ($\diamond$) are from \cite{Gibs67a}.} \vspace{-3mm}
\end{minipage} 
\hspace{5mm}
\begin{minipage}[t]{7.5cm}
  \epsfig{file=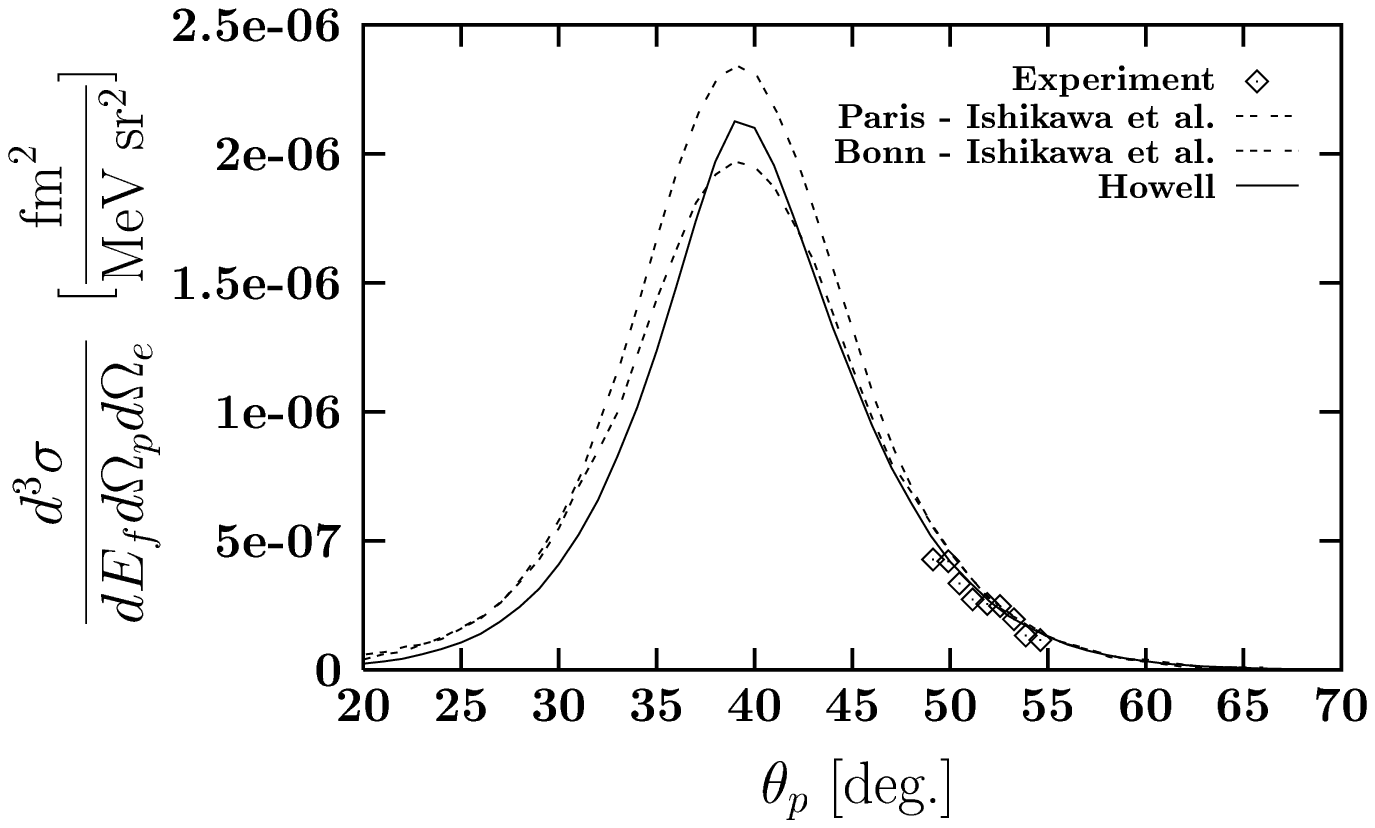,width=7.5cm,height=5.2cm} \vspace{-14mm}
\caption{\label{figelectro2} Coincidence cross section
  for the C2 kinematics of \cite{Ishi94a}.} \vspace{-3mm}
\end{minipage}
\end{figure}

Early calculations of electron-induced nuclear disintegration processes, with
the scattered electron detected in coincidence with the ejected proton
\protect{(''coincidence cross section'')}, were performed by Griffy and Oakes
\cite{Griff64a} using the Irving three-body wave functions or a pole
approximation approach. The corresponding results (dashed or dotted lines) are
shown in Fig. \ref{figelectro1}.  Few-body techniques were employed by Lehman
to obtain more appropriate wave functions. He calculated in this way the two-
and three-fragment electrodisintegration of $^{3}$H and $^{3}$He for the
Yamaguchi potential, although without FSI \cite{Lehm69a}.

The Gibson-Lehman approach, i.e., the Faddeev-AGS-type treatment of
photodisintegration \cite{Gibs75} discussed in Sec.~1.2 was extended to
$^{3}$H and $^{3}$He electrodisintegration by Heimbach et~al.
\cite{Heimb77a}.  Being based on the fairly simple Yamaguchi potential, this
procedure takes into account consistently the final-state interaction. Van
Meijgaard and Tjon \cite{Meijgaard90a} employed the MT I+III potential in
unitary pole expansion (UPE), and found a considerable difference between the
Yamaguchi and MT results.

More realistic interactions, the Urbana and Argonne potentials, were used in
the variational Monto Carlo (VMC) \cite{Schiavilla86a} calculations by
Schiavilla and Pandharipande \cite{Schiavilla87a}, while Laget \cite{Laget85a}
employed a diagrammatic expansion of the amplitudes, taking into account in
this way at least part of FSI and MEC.

In a more rigorous approach Ishikawa et~at.~\cite{Ishi94a} calculated $^{3}$He
electrodisintegration by a 34-channel treatment of Faddeev-type
two-dimensional equations.  The results for the Paris- and Bonn {\sl B}
potentials obtained in this way are shown in Fig.  \ref{figelectro2}.

\subsection{Four-nucleon problem}

In the computationally much more demanding electrodisintegration of $^{4}$He
into $p + \,^{3}$H or $n + \,^{3}$He practically no exact calculations of the
final $^{(-)}\langle \vec q^{\, \prime}; \psi_{III}|$ scattering state, and
thus of FSI, has been performed. One therefore resorts to the approximation

\begin{equation}
\label{eqapprox}
^{(-)}\langle \vec q^{\, \prime}; \psi_{A-1}|
\simeq \, {^{(-)}}\langle \vec q^{\, \prime}| \langle \psi_{A-1}| \,,
\end{equation}

\noindent
traditionally used in nuclear physics.  Here, $^{(-)}\langle \vec
q^{\,\prime}|$ represents the scattering state of the elementary particle
interacting with the composite particle $\langle \psi_{A-1}|$ via an
appropriately chosen optical potential. A decisive criterion for the quality
of this approximation is that it does not violate the orthogonality condition
between bound and scattering states.  In other words, the condition

\begin{equation}
\label{eqortho}
^{(-)}\langle \vec q^{\, \prime}| \, \langle \Psi_{A-1}|\Psi_{A} \rangle
\simeq 0 
\end{equation}

\noindent
has to be satisfied with high accuracy. We refer in this context to the
investigations by Schiavilla and Pandharipande \cite{Schiavilla87a} and to the
extension of their above-mentioned three-body calculations to the four-body
case \cite{Schiavilla90a}.  The three-body approach by Laget has also its
four-body counterpart \cite{Laget94a}.  The main findings of these
investigations were that in certain kinematical regions there is an extreme
sensitivity to the nuclear forces and that the different models employed lead
to fairly distinct results.

\subsection{Marchenko inversion approach}

The considerable effort required when treating three- and four-body
disintegration processes in their full complexity, suggests to look more
closely at approximations of the type (\ref{eqapprox}). Assuming this
approximation to be justified, the main problem consists in constructing the
effective interaction between the outgoing fragments in a reliable manner.

L.L. Howell et~al. \cite{Howellcontribution,Howellsubmitted} employed for this
purpose the Marchenko inversion method. The bound states are calculated by
means of the comparatively simple integrodifferential equation approach
(IDEA) \cite{Ripelle88}.  The,
in principle, exact Marchenko treatment requires the knowledge 
of the high-energy behavior of the nucleon-nucleus phase shifts. The lack of
data at such energies allows for various extrapolations leading to different
effective potentials and thus, via the corresponding scattering states
$^{(-)}\langle \vec q \,|$, to different electrodisintegration results. In
other words, proceeding in this way provides an appropriate tool for testing
the high energy behavior of the phase shifts in a fairly sensitive physical
situation.

\begin{figure}[htb]
  \centerline{ \epsfig{file=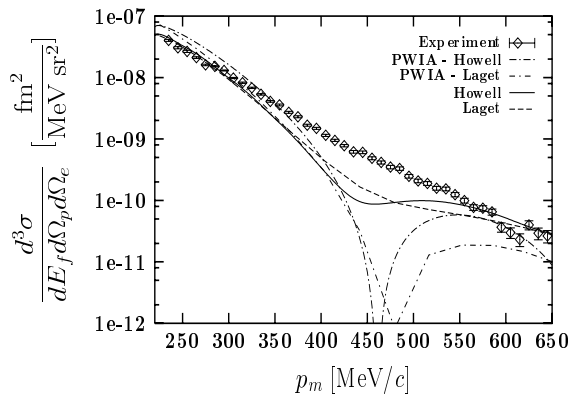,width=7.5cm,height=5.5cm}} \vspace{-8mm}
\caption{\label{figelectro3} Coincidence cross section for the
  two-fragment electrodisintegration of $^4$He as a function of the missing
  momentum for the kinematics (A,B,C,D) of Ref.~\cite{Leeuwe96a}.}
\vspace{-3mm}
\end{figure}

The apparent freedom in extrapolating the phase shifts practically vanishes
when imposing the orthogonality condition (\ref{eqortho}) on the resulting
wave function $^{(-)}\langle \vec q^{\, \prime}|$.  The solid line in Fig.
\ref{figelectro1} obtained in this way agrees remarkably well with the
pole-approximation results and thus with the data.  In Fig. \ref{figelectro2}
the corresponding solid line lies between the Paris and Bonn results of
Ishikawa et~al. \cite{Ishi94a}.  In other words, this model provides highly
reliable results within a comparatively simple model approach, which moreover
sheds some light on the high-energy behavior of the nucleon-nucleus phase
shifts.  It, hence, leads to quite relevant informations concerning the
construction of optical potentials.

An example for $^{4}$He electrodisintegration at higher energies is shown in
Fig. \ref{figelectro3} The characteristic dip found in the plane wave (Born)
approximation is partly, but not fully, removed when taking into account the
FSI by means of Laget's diagrammatic technique \cite{Laget94a}. The Marchenko
inversion approach by Howell et~al. \cite{Howellcontribution,Howellsubmitted}
agrees fairly well with the corresponding results.  That is, both in the three-
and in the four-body case it represents a rather natural, comparatively simple
alternative to other methods.

With respect to the remaining discrepancies in the dip region, which are
neither cured by the Laget nor the Marchenko inversion procedure, it appears
necessary to incorporate, at least approximately, the interplay of the (3 + 1)
and (2 + 2) amplitudes , which appears quite essential in the fully coupled
system of equations. (compare in this context the discussion in Sec.~2.1).
Investigations in this direction are in progress.

%\bibliography{physik} \bibliographystyle{prsty}

%\end{document}

\end{document}